\documentclass[sigconf]{acmart}

\usepackage{float}
\usepackage{booktabs}
\usepackage{subcaption}
\usepackage{listings}
\usepackage{graphicx}

\title[Machine Learning Detection of Telegram Pump-and-Dump Schemes]{Real-Time Machine Learning Detection of\\Telegram-Based Pump-and-Dump Schemes}

\author{Manuel Bolz}
\affiliation{%
  \institution{University of Zurich}
  \country{Switzerland}
}

\author{Kevin Bründler}
\affiliation{%
  \institution{University of Zurich}
  \country{Switzerland}
}

\author{Liam Kane}
\affiliation{%
  \institution{University of Zurich}
  \country{Switzerland}
}

\author{Panagiotis Patsias}
\affiliation{%
  \institution{University of Zurich}
  \country{Switzerland}
}

\author{Liam Tessendorf}
\affiliation{%
  \institution{University of Zurich}
  \country{Switzerland}
}

\author{Krzysztof M. Gogol}
\authornote{Corresponding author: \href{mailto:gogol@ifi.uzh.ch}{gogol@ifi.uzh.ch}}
\affiliation{%
  \institution{University of Zurich}
  \country{Switzerland}
}

\author{Taehoon Kim}
\affiliation{%
  \institution{University of Zurich}
  \country{Switzerland}
}

\author{Claudio J. Tessone}
\affiliation{%
  \institution{University of Zurich}
  \country{Switzerland}
}

\begin{document}

\begin{abstract}
Cryptocurrency markets often face manipulation through prevalent pump-and-dump (P\&D) schemes, where self-organized Telegram groups, some exceeding two million members, artificially inflate target cryptocurrency prices. These groups sell premium access to inside information, worsening information asymmetry and financial risks for subscribers and all investors. This paper presents a real-time prediction pipeline to forecast target coins and alert investors to possible P\&D schemes. In a Poloniex case study, the model accurately identified the target coin among the top five from 50 random coins in 24 out of 43 (55.81\%) P\&D events. The pipeline uses advanced natural language processing (NLP) to classify Telegram messages, identifying 2,079 past pump events and detecting new ones in real-time.
\end{abstract}




\acmConference[DeFi'25]{ACM CCS Workshop on Decentralized Finance and Security}{October 17, 2025}{Taipei, Taiwan}
\acmBooktitle{Proceedings of the ACM CCS Workshop on Decentralized Finance and Security (DeFi'25), 
October 17, 2025, Taipei, Taiwan}

\maketitle

\section{Introduction}
\label{sec:Introduction}

Blockchain is an immutable, decentralized ledger that eliminates the need for a central authority, enabling secure and transparent transactions. It facilitates the continuous tracking of ownership through digital tokens, which act as units of value. Tokens can store value, serve as participation rewards, provide voting rights, or enable interaction with products and services \cite{pilkington2016blockchain, 9219709, gogol2024sokdefi}. However, tokens are also susceptible to market manipulations, making their attributes critical to this study.

Cryptocurrency markets promise high returns and decentralized control but remain vulnerable to market manipulations like pump-and-dump (P\&D) schemes. These schemes, orchestrated via messaging platforms such as Telegram and Discord, artificially inflate a cryptocurrency's price, enabling organizers to profit at the expense of unsuspecting investors. A recent report by Chainalysis \cite{chainalysis2023crypto} revealed that 24\% of tokens launched in 2022 exhibited characteristics typical of P\&D schemes, underlining the systemic risks these activities pose to market integrity and the urgent need for effective detection and prevention mechanisms.

P\&D schemes typically unfold in distinct phases: an initial announcement, a countdown to build anticipation and attract participants, and the final release of the target coin, which sparks a rapid price surge as traders rush to buy. This is immediately followed by a sell-off, where organizers and early participants offload their holdings at inflated prices, leaving latecomers with significant losses. The cycle often repeats, exploiting market vulnerabilities and undermining trust among investors, as depicted in the list below.
\begin{enumerate}
    \item \emph{Pump Announcement:} Formal announcements detailing the exchange, time, and date of the event.
    \item \emph{Countdown:} Reminders and promotional messages leading up to the pump event.
    \item \emph{Target Coin Release:} Disclosure of the target coin's symbol, token contract, or trading pair link, often as text or images.
    \item \emph{Pump Results:} Summaries of the pump results, including profits and performance.
    \item \emph{Delay or Cancellation Notices:} Updates about postponed or canceled pumps.
    \item \emph{Noise:} Messages unrelated to the categories above.
\end{enumerate}

Although traditional analysis of trading volume and token price can provide valuable indicators for the emergence of P\&D schemes~\cite{kamps2018moon}, recent advances in large language models (LLMs)~\cite{2024SurveyGTP3,2024SurveyLLM,wang2024survey} allow for better detection and understanding of these manipulations. LLMs enable detailed analysis of Telegram messages, often the primary medium through which P\&D organizers communicate with participants. By extracting patterns and identifying key linguistic cues in these messages, LLMs can complement conventional market data analysis and offer a more comprehensive detection approach.

Additionally, the execution of P\&D schemes on centralized cryptocurrency exchanges requires a sufficiently liquid and deep order book to accommodate the influx of trades and facilitate the artificial price rise. Analyzing order book depth and anomalies can provide critical insights into the preparatory activities of orchestrators, including potential insider trades and coordinated buy orders.

The introduction of new token standards, such as Ordinals~\cite{rodarmor-2023}, Inscriptions~\cite{li-2024}, and Runes~\cite{rodarmor2023runes, vermaak2023runes, kucoin2024runes}, on Bitcoin, further complicates the landscape of P\&D schemes. These standards spread to the EVM-chains~\cite{messias2024Inscription}, raising questions about the susceptibility of tokens beyond the widely targeted ERC-20 standard on Ethereum and other EVM-compatible chains. While ERC-20 tokens have historically been prime targets due to their widespread adoption and ease of trading, understanding whether and how emerging standards are exploited is crucial for developing holistic anti-manipulation strategies.

By integrating advanced NLP techniques, market data analysis, and an understanding of evolving token standards, this study provides new contributions to the detection and prevention of P\&D schemes, offering a novel framework for safeguarding cryptocurrency markets against manipulation.

\subsection*{Related Work}
Early studies on P\&D schemes  focused on identifying price and volume anomalies after pumps occurred \cite{kamps2018moon}, achieving moderate success but underscoring the need for more sophisticated models. Real-time detection approaches, such as anomaly detection using machine learning \cite{mansourifar2020hybrid}, have shown promise but suffer from latency issues, as seen in the 30-minute lag of certain methods \cite{victor2019cryptocurrency}. More recent methods \cite{la2023doge} leverage Random Forest and AdaBoost classifiers to detect anomalies within seconds of pump initiation.

Target coin prediction aims to identify manipulated coins before a pump. Xu et al. \cite{xu2019anatomy} developed Random Forest models to predict pump likelihood using market metrics, while Hu et al. \cite{hu2023sequence} introduced sequence-based deep-learning models leveraging channel-specific features. Despite advancements, existing approaches largely rely on historical data and lack integration of high-frequency order book data, limiting their real-time applicability.

\subsection*{Contributions}
This paper addresses the limitations of existing methods by introducing a cross-exchange, real-time pipeline for detecting P\&D schemes before they occur. Our contributions include:
\begin{enumerate}
    \item Incorporating high-frequency order book and trade data alongside Telegram messages monitoring to enhance prediction accuracy.
    \item Developing a Z-score-based model, which can forecast target coins mere seconds prior to pump events, achieving correct predictions among the top five ranked coins in 55.81\% of cases and within the top ten in 74.42\% of instances.
    \item By providing an early warning system, this framework aims to mitigate market manipulation risks and promote safer trading environments.
\end{enumerate}

\subsection*{Paper Organization}
The paper is structured as follows: Section~\ref{sec:Background} introduces the relevant token standards. Section \ref{sec:System} presents the pipeline system architecture for real-time prediction of P\&D schemes, Section~\ref{sec:Data} details the data collection process. Empirical results are presented in Section~\ref{sec:Results}, followed by a discussion in Section~\ref{sec:Discussion}, and conclusions in Section~\ref{sec:Conclusion}.

\section{Preliminaries}
\label{sec:Background}

To provide context for this research, we briefly introduce main token standards. 


\noindent
\textbf{ERC-20.}
ERC-20 is a widely adopted standard for fungible tokens on Ethereum, enabling interoperability across decentralized applications (dApps). Tokens under this standard can represent anything from assets to utility tokens, facilitating broad use cases \cite{morales2023erc-20, ERC20}. BEP-20, a similar standard on Binance Smart Chain, offers lower transaction fees and congestion, enhancing its appeal \cite{alphapoint2024erc20vsbep20}.

\noindent
\textbf{ERC-721.}
ERC-721 defines a standard for non-fungible tokens (NFTs), which are unique and indivisible. NFTs enable the representation of ownership for digital assets like art and collectibles, ensuring their provenance and authenticity through blockchain's immutable records \cite{barrington2022erc-721, ERC721}.

\noindent
\textbf{Ordinals and Inscriptions.}
Ordinals enable individual satoshis to carry unique identifiers, granting them subjective value based on collector sentiment \cite{rodarmor-2023}. Inscriptions enhance this concept by attaching data like images or text to satoshis, transforming them into unique digital artifacts that can be traded \cite{Inscriptions}.

\noindent
\textbf{BRC-20.}
The BRC-20 standard introduces fungible tokens on Bitcoin, leveraging the Ordinals protocol to inscribe metadata onto satoshis. While BRC-20 enables token creation and transfer, its programmability is limited compared to Ethereum-based standards \cite{li-2024}. This simplicity highlights its experimental nature, relying on off-chain tools for effective tracking and management.

\noindent
\textbf{Runes.}
Runes embed fungible tokens directly into Bitcoin's UTXO system, allowing for efficient token creation, minting, and transfer. The protocol ensures integrity through a burning mechanism for invalid transactions and maintains a smaller on-chain footprint, addressing scalability concerns \cite{rodarmor2023runes, vermaak2023runes, kucoin2024runes}.


\section{Pipeline System Architecture}
\label{sec:System}

\begin{figure*}[tb]
\centering
\includegraphics[width=0.8\textwidth]{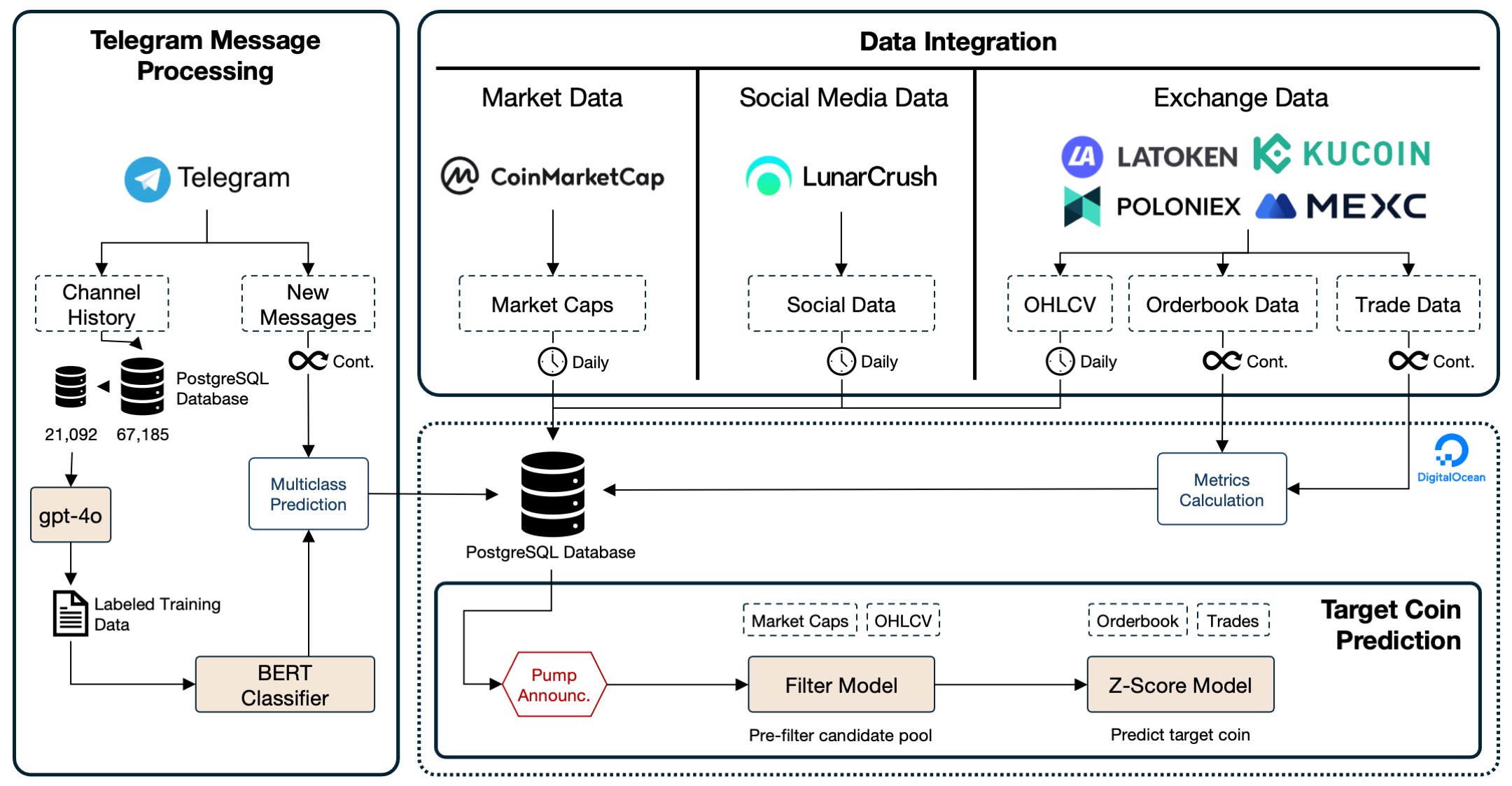}
\caption{Architecture of the Real-time P\&D Detection Pipeline.}
\label{fig:real_time_pipeline}
\end{figure*}

Figure \ref{fig:real_time_pipeline} illustrates the architecture of our detection pipeline system that is designed to monitor, integrate and analyze data streams in real-time to detect P\&D schemes effectively. This pipeline is fully operational and deployed on the cloud infrastructure. It comprises three main components.

\paragraph*{Telegram Message Processing} The system actively monitors Telegram channels, continuously classifying incoming messages into six distinct categories that correspond to different phases of P\&D schemes. By employing advanced natural language processing (NLP) techniques, the model efficiently identifies announcements related to upcoming events. These classified messages are stored in a PostgreSQL database, serving as the primary trigger for subsequent stages of the pipeline.

\paragraph*{Data Integration} To complement the Telegram data, the pipeline incorporates high-frequency market data, including order book and trade data, sourced from multiple exchanges such as LATOKEN, KuCoin, MEXC, Poloniex, and XT.com. Covering approximately 4,400 coins, this module processes data streams alongside market indicators to compute essential metrics. These metrics enable the pipeline to track market movements and identify suspicious patterns in real-time, ensuring comprehensive and dynamic monitoring.

\paragraph*{Target Coin Prediction} The final stage of the pipeline predicts the coin most likely to be targeted in a P\&D scheme. This is achieved through a two-step process. Initially, the set of candidate coins is narrowed down using a filter based on historical market capitalization data. Subsequently, a statistical anomaly detection model, leveraging Z-scores, calculates the likelihood of each coin being pumped. This process generates a ranked list of potential target coins. In a case study involving 43 pump events on Poloniex, the system accurately identified the pumped coin within the top five predictions in 55.81\% of cases (24 out of 43 events), showcasing the pipeline's precision and real-time effectiveness.

The detection pipeline system is hosted on the cloud infrastructure, leveraging multiple virtual machines with 8 GB of memory and 160 GB of disk space to handle high-frequency processing from multiple cryptocurrency exchanges. Data is stored in a PostgreSQL database, which, as of October, 2024, contains over 91,295 labeled and processed Telegram messages. This database is continuously updated to reflect ongoing P\&D schemes.
\section{Data Collection and Methodology}
\label{sec:Data}


\begin{table}[htbp]
    \centering
    \caption{Summary of data sources utilized in P\&D detection pipeline, including purpose, scope, and relevant details.}
    \label{tab:datasources}
    \begin{tabular}{p{2.8cm}|p{4.5cm}}
        \hline
        \textbf{Data Source} & \textbf{Details} \\ 
        \hline
        TGstats & 43 active Telegram channels identified for P\&D events. \\ 
        \hline
        Telegram & 91,295 P\&D messages (growing); data from 2017-12-02 to 2024-10-21. \\ 
        \hline
        LunarCrush & Social media metrics for cryptocurrencies. \\ 
        \hline
        CoinMarketCap & Market caps and token metadata. \\ 
        \hline
        CoinCodex & Historical market caps for 3,958 unpumped and 924 pumped coins (2018-01-01 to 2024-07-01). \\ 
        \hline
        CEXs (e.g., KuCoin, Poloniex) & Daily OHLCV data for 4,643 cryptocurrencies (365,982 rows) from 2024-07-01 to 2024-10-21. \\ 
        \hline
        CEXs (e.g., KuCoin, Poloniex) & Order book data (5–44M rows) and trade data (2–5M rows per exchange) from 2024-07-01 to 2024-10-21. \\ 
        \hline
    \end{tabular}
\end{table}

The pipeline incorporates 365,982 rows of daily OHLCV (open, high, low, close, volume) price data, social metrics, and market capitalization data for 4,643 cryptocurrencies. Additionally, it stores between 5 and 44 million rows of order book metrics and between 2 and 5 million rows of trade metrics per exchange. These metrics, crucial for target coin prediction, are calculated and updated every few seconds. Table \ref{tab:datasources} presents a comprehensive overview of these data sources, which are elaborated upon in this section. Data was collected from 30 days up to 1 hour before each pump event.

\subsection{Pump Message Collection}
\label{sec:Messages}
Previous studies on target coin prediction \cite{xu2019anatomy} primarily relied on data from the now-defunct PumpOlymp\footnote{PumpOlymp was previously active under \url{https://pumpolymp.com/}} website to identify pump channels on Telegram. In the absence of a comparable successor, we utilized TGStats \cite{tgstat} to search for specific keywords within channel titles and descriptions, including \textit{crypto}, \textit{pump}, \textit{dump}, \textit{P\&D}, and \textit{signal}, along with their combinations. Identified channels were manually validated for active promotion of P\&D schemes, resulting in the identification of 43 active channels. From these, we extracted complete message histories, creating a dataset currently comprising 91,295 messages. Telegram was chosen as the primary focus, as previous studies have established its role as the main platform for P\&D activities \cite{xu2019anatomy}.

\subsection{Pump Message Labeling}
Manual examination of the collected messages revealed six distinct categories of P\&D-related messages, aligned with the pump anatomy described in Section \ref{sec:Introduction}.
The dataset was preprocessed to normalize characters, spaces, and emojis. User-tags were replaced with the generic token \texttt{@USER}, and non-informative URLs were replaced with \texttt{UNKNOWN\_URL}. A list of prominent centralized exchanges (CEXs), decentralized exchanges (DEXs), and defunct platforms such as HotBit and Cryptopia was compiled to tag messages mentioning exchanges with special tokens (\texttt{\_CEX} or \texttt{\_DEX}).

To automate P\&D detection, we labeled approximately 25\% (21,092 messages) of the dataset across the six categories using the GPT-4o model via the OpenAI API. Detailed prompts and parameter settings are available in Appendix \ref{app:Prompts}. Manual verification and re-labeling were performed to enhance label accuracy, focusing on critical classes: Pump Announcement, Countdown, Target Coin Release, and Delay/Cancellation. Only a small number of messages required re-labeling, demonstrating the reliability of GPT-4o as a baseline model.

\begin{table}[h!]
    \centering
    \begin{minipage}[t]{0.48\textwidth}
        \centering
        \caption{Distribution of labels assigned to Telegram messages.}
        \label{tab:training_data}
        \begin{tabular}{lr}
        \toprule
        \textbf{Label} & \textbf{Count} \\
        \midrule
        Pump Announcement & 1,184 \\
        Countdown & 11,321 \\
        Target Coin Release & 1,151 \\
        Pump Results & 1,722 \\
        Delay/Cancellation & 98 \\
        Noise & 5,616 \\
        \midrule
        \textbf{Total} & \textbf{21,092} \\
        \bottomrule
        \end{tabular}
    \end{minipage}%
    \hfill
    \begin{minipage}[t]{0.48\textwidth}
        \centering
        \caption{PumpBERT classification performance.}
        \label{tab:model_performance}
        \begin{tabular}{lr}
        \toprule
        \textbf{Metric} & \textbf{Value} \\
        \midrule
        \textbf{Overall Metrics} & \\
        \midrule
        F1 weighted average & 0.982 \\
        Precision & 0.982 \\
        Recall & 0.982 \\
        \midrule
        \textbf{Label-specific F1 Scores} & \\
        \midrule
        Pump Announcement & 0.976 \\
        Countdown & 0.993 \\
        Target Coin Release & 0.995 \\
        Pump Results & 0.970 \\
        Delay/Cancellation & 1.000 \\
        Noise & 0.971 \\
        \bottomrule
        \end{tabular}
    \end{minipage}
\end{table}

The labeled data was used to fine-tune a BERTweet \cite{nguyen2020bertweetpretrainedlanguagemodel} model. Table \ref{tab:training_data} summarizes the label distribution after cleaning. To address label imbalance, a weighted loss function was applied during training. The dataset was split into 80\% for training and 20\% for validation and testing. Duplicate messages, commonly reused across channels (e.g., "1 HOUR UNTIL THE PUMP..."), were excluded from the test set to ensure unbiased evaluation metrics. The performance of the model is included in Table \ref{tab:model_performance}.

The model was then used to predict labels for the remaining messages. From the labeled message history, we identified 2,079 distinct pump events by clustering messages based on token symbol, exchange, and the timestamp of the Target Coin Release, rounded to the nearest hour. For channels disclosing target coins via images, we manually extracted the corresponding symbols. This occurred for only three channels in the sample.

\subsection{Market Data} We analyze 365,982 rows of daily OHLCV data for 4,643 cryptocurrencies, including price, trading volume, and market capitalization metrics critical for identifying P\&D targets. Data is collected via API calls to exchanges such as KuCoin and MEXC, supplemented by the CCXT library \cite{ccxt}, and automated through a daily cron job. Preprocessed data is stored in a database and used in filter models for predictive analysis.

Challenges in historical data accuracy arose due to non-unique coin symbols and missing data for 403 of the 1,045 unique symbols, often due to delistings or defunct exchanges. To address this, our real-time pipeline incorporates self-reported market capitalization data from CoinMarketCap~\cite{CoinMarketCap}.
Additionally, derivative tokens, such as leveraged and inverse tokens, were excluded using a pattern-matching approach, ensuring that only primary cryptocurrencies were considered.

\subsection{Social Data} Social sentiment significantly impacts cryptocurrency prices \cite{wolk2019}. Using the LunarCrush API \cite{lunarcrush2024}, we collect metrics such as likes, shares, and social dominance. Features such as rolling interaction means and social dominance ratios are engineered to explore correlations with market trends. 

\subsection{Order Book Data} Real-time order book data offers early indicators of market manipulation. Metrics such as bid-ask spread, order size, and imbalance ratios are calculated from WebSocket feeds focused on USDT pairs flagged as vulnerable to P\&D schemes. 

\subsection{Trade Data} Trade data provides a historical record of executed transactions, revealing actual market behavior. Metrics such as trade volume, VWAP, and taker side volume help differentiate buy and sell pressures. 
Integrating trade data enhances the pipeline's predictive accuracy by capturing a complete view of market activity.

\subsection{Target Coin Normalization}
\label{sec:Z-Score}

Our Z-score-based statistical model leverages historical and short-term data from order book and trade metrics to identify market configurations indicative of impending P\&D schemes. By analyzing patterns such as unusual sell orders or anomalous buy trades, the model aims to detect signs of organizers accumulating positions prior to a pump and placing sell orders to realize profits during the event.

The model processes data from a three-day window leading up to each pump event for all filtered coins on the exchange. Key metrics include order book pressure, average order size, order imbalance ratio, market order impact, volume-weighted average price (VWAP), high-low spread, and trade count. For each coin, Z-scores are calculated to compare short-term market behavior against historical averages, quantifying deviations from usual activity. The Z-score formula is:

\[
z_i = \frac{x_i - \mu}{\sigma}
\]

where \(z_i\) is the Z-score of the \(i\)-th observation, \(x_i\) is the short-term metric value, \(\mu\) is the historical mean, and \(\sigma\) is the historical standard deviation. The resulting Z-scores are normalized and ranked, providing an estimate of a coin being targeted for a pump.

\paragraph*{Backtesting and Results}
We evaluated the model on 43 P\&D events using trade and order book data from the Poloniex API. For each event, the dataset included data for the pumped coin and a random sample of 50 coins. Figure~\ref{fig:histogram_rank_pummps} illustrates the distribution of ranks for pumped coins based on Z-scores. When the model was run 20 seconds before the pump, the pumped coin ranked within the top five (TOP5) in 55.81\% of cases and within the top 10 (TOP10) in 74.42\%. Using only trade data, these percentages dropped to 46.51\% and 55.81\%, respectively. For order book data alone, the TOP5 and TOP10 performances were 44.19\% and 72.09\%.

\begin{figure}[htpb]
    \centering
    \includegraphics[width=0.99\linewidth]{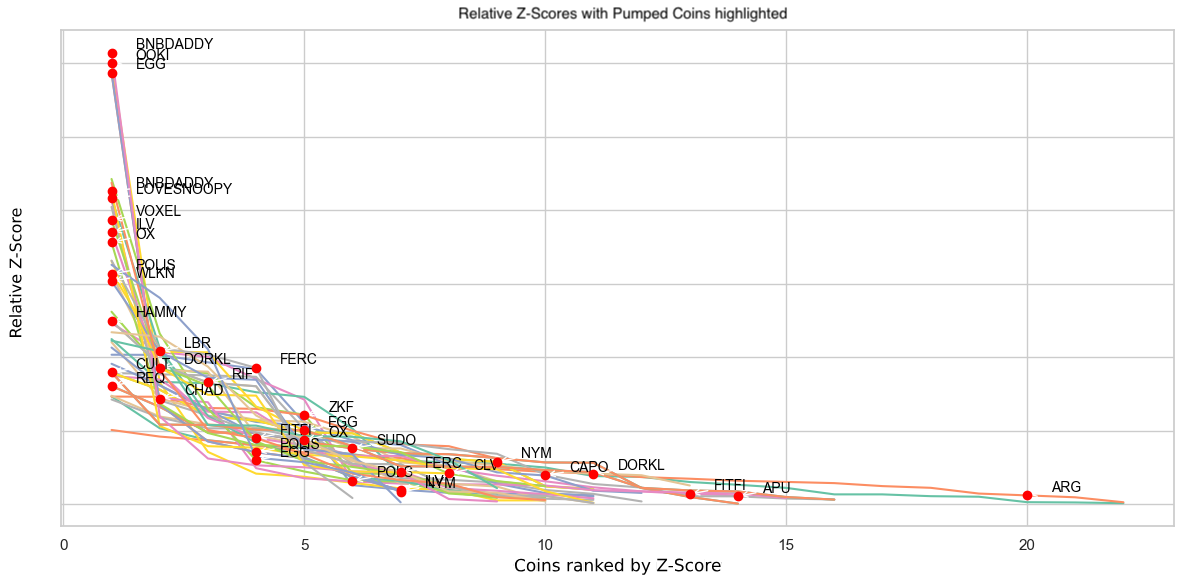}
    \caption{Relative Z-Scores with Pumped Coins Highlighted}
    \label{fig:eval_pd_pumps}
\end{figure}

Figure~\ref{fig:eval_pd_pumps} visualizes the relative Z-scores across all coins for each pump event, with the pumped coins highlighted. This demonstrates the model’s ability to distinguish pumped coins based on their Z-scores.

\paragraph*{Performance Over Different Time Offsets}
We tested the model at time offsets of 20 seconds, 40 seconds, and 1 minute before the pump. As shown in Table~\ref{tab:TOP5}, TOP5 and TOP10 performances declined with increasing time offsets, highlighting the importance of near-real-time predictions. When using both trade and order book data, the TOP5 performance decreased from 55.81\% (20 seconds) to 19.05\% (1 minute). Similarly, using only trade data, TOP5 performance dropped from 46.51\% to 9.52\%.

\begin{table}[htbp]
    \centering
    \caption{TOP5 and TOP10 Performance with Different Time Offsets}
    \label{tab:TOP5}
    \resizebox{\columnwidth}{!}{%
    \begin{tabular}{l|r r|r r|r r}
        \toprule
        \textbf{Time Offset} & \multicolumn{2}{c|}{\textbf{Trade and Order Book Data}} & \multicolumn{2}{c|}{\textbf{Trade Data Only}} & \multicolumn{2}{c}{\textbf{Order Book Data Only}} \\
         & TOP5 & TOP10 & TOP5 & TOP10 & TOP5 & TOP10 \\
        \midrule
        20 seconds & 55.81\% & 74.42\% & 46.51\% & 55.81\% & 44.19\% & 72.09\% \\
        40 seconds & 41.46\% & 60.98\% & 29.27\% & 36.59\% & 41.46\% & 60.98\% \\
        1 minute & 19.05\% & 28.57\% & 9.52\% & 11.90\% & 16.67\% & 30.95\% \\
        \bottomrule
    \end{tabular}
    }
\end{table}

The Z-score model effectively identifies target coins shortly before P\&D events, with combined trade and order book data yielding the best results. However, its predictive power diminishes as the time offset increases, emphasizing the need for real-time data processing in operational pipelines.
\section{Empirical Results}
\label{sec:Results}

\begin{figure*}[h]
\centering
\includegraphics[width=0.7\textwidth]{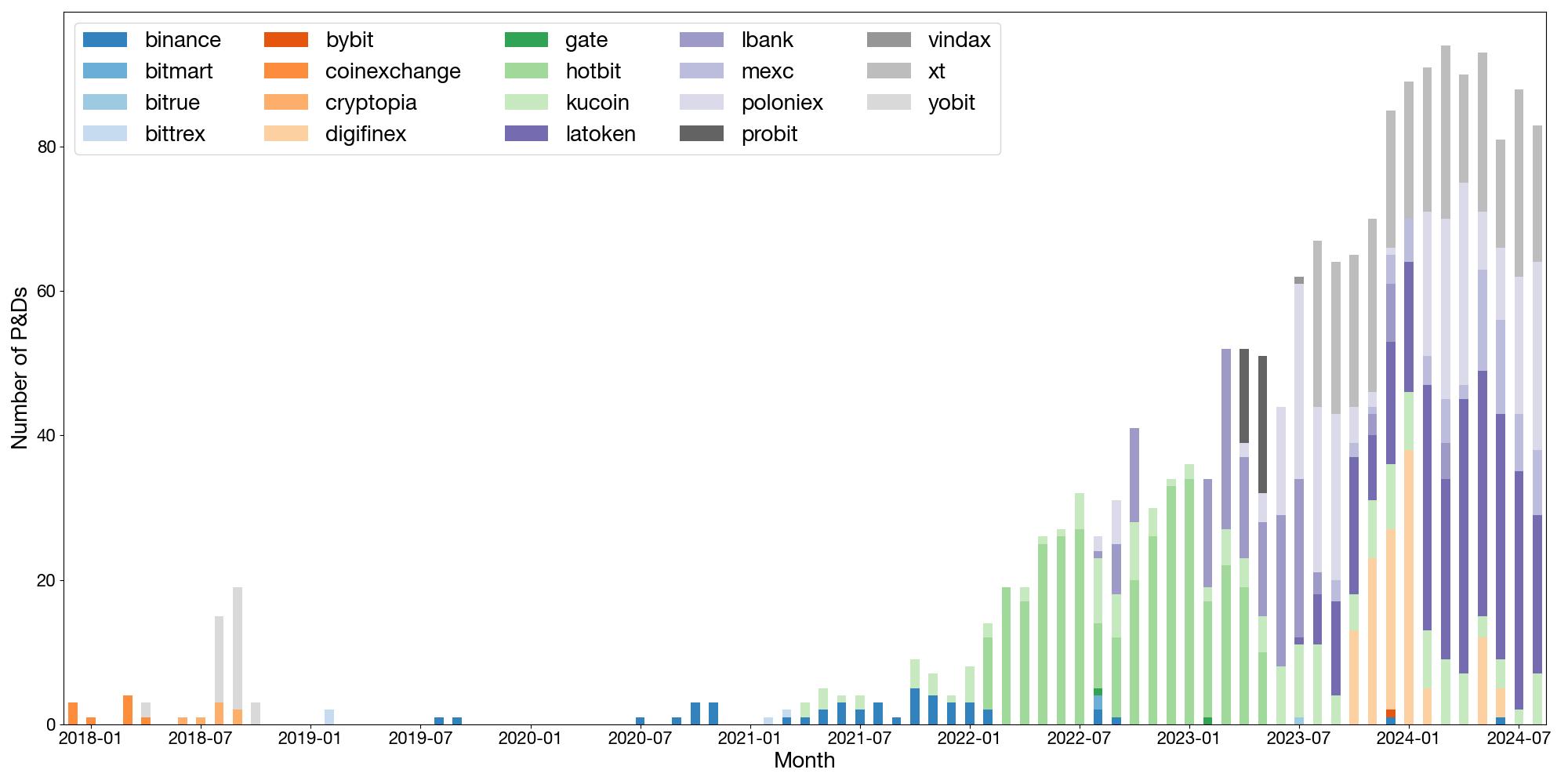}
\caption{Temporal Distribution of P\&D Events by Exchange.}
\label{fig:stacked_bar_plot}
\end{figure*}

First, the model labeled messages and extracted details such as pump times, exchanges, and target coins. This yielded a dataset of 2,079 P\&D events between December 2017 and September 2024. As shown in Table~\ref{tab:pumps_by_exchange} in Appendix, most events occurred on centralized exchanges (CEX), with Hotbit leading despite ceasing operations in May 2023. Recent activity has shifted to LATOKEN, XT, and Poloniex, as illustrated in Figure~\ref{fig:stacked_bar_plot}.

\subsection{Quantification of P\&D Schemes and Target Coins}

\begin{figure*}[h]
\centering
\includegraphics[width=0.7\textwidth]{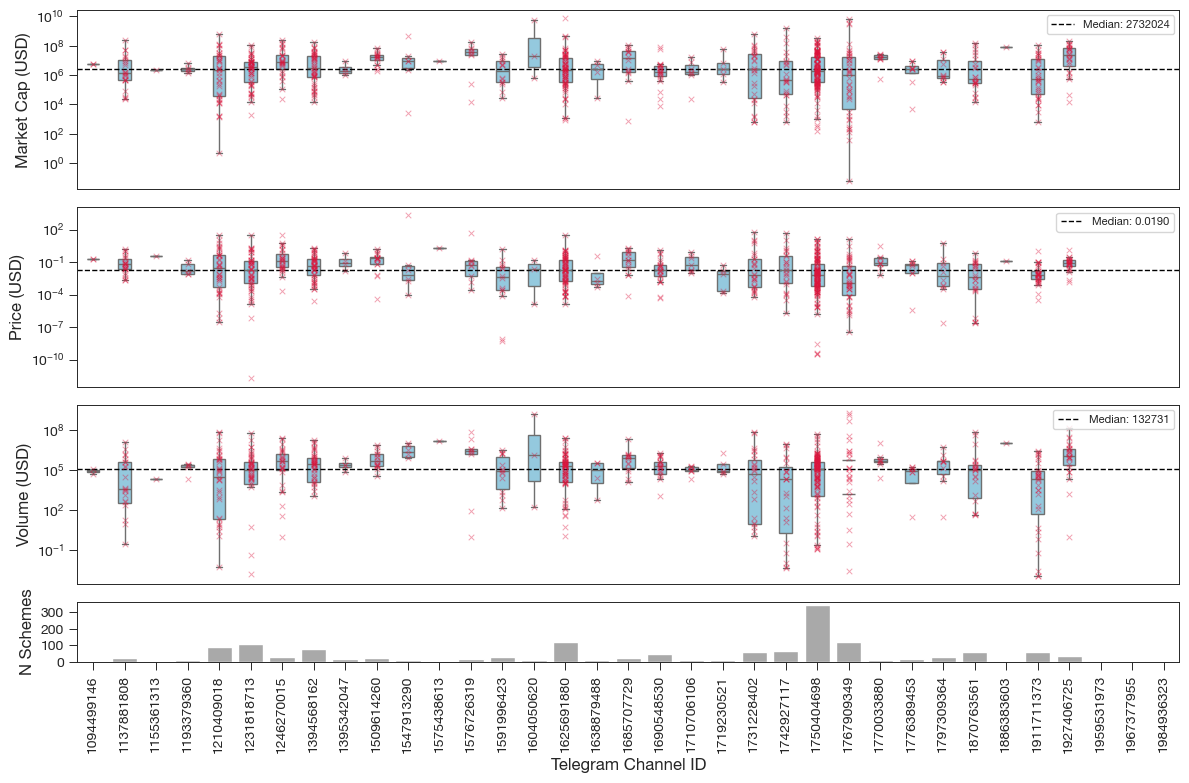}
\caption{Market Caps, Price, and Volume Distribution of Pumped Coins per Channel}
\label{fig:mcap_price}
\end{figure*}

We analyzed market cap and price data from CoinCodex~\cite{CoinCodex} for pumped coins. Supplementary Figure~\ref{fig:mcap_price} in Appendix highlights that most pumped coins belong to the small- and mid-cap categories. 
Many coins were targeted multiple times, with some becoming frequent targets. Table~\ref{tab:top_5} in Appendix lists the five most pumped coins, led by TOKKI, which was pumped nine times across four exchanges by five different channels.

\subsubsection{Volume}

Our analysis reveals a significant impact of pump events on the trade volumes of targeted coins. Supplementary Figures~\ref{fig:avg_daily_volumes_kucoin} and~\ref{fig:avg_daily_volumes_poloniex} in Appendix illustrate the average daily trade volumes for approximately 700 pump events on KuCoin and Poloniex over an eight-day period, with the eighth day representing the pump event. The substantial increase in trade volume on the pump day compared to a typical trading day is strikingly evident.

While KuCoin exhibits higher absolute volumes on the pump day (approximately \$1.86 million) compared to Poloniex (\$4,800), the relative increase in trade volume is greater on Poloniex. Specifically, the ratio of the pump day volume to the day before the pump is approximately \(62.5\) for Poloniex (\(4809 / 77\)) versus \(23.2\) for KuCoin (\(1867411 / 80641\)), suggesting that pump events have a more pronounced relative impact on smaller exchanges.

We further analyzed average trade volumes grouped by the Telegram channels orchestrating these pump events. Supplementary Figures~\ref{fig:avg_daily_volumes_by_channels_kucoin} and~\ref{fig:avg_daily_volumes_by_channels_poloniex} in Appendix highlight the variation in trade volume increases across different channels on the pump day.
Interestingly, the observed variations cannot be solely attributed to easily measurable factors such as the number of subscribers per channel. This indicates that additional, less apparent factors play a role in determining the magnitude of a pump event's impact.

\subsubsection{Price Spikes}

\begin{figure*}[thb]
    \centering
    \begin{minipage}[t]{0.48\textwidth}
        \centering
        \includegraphics[width=\textwidth]{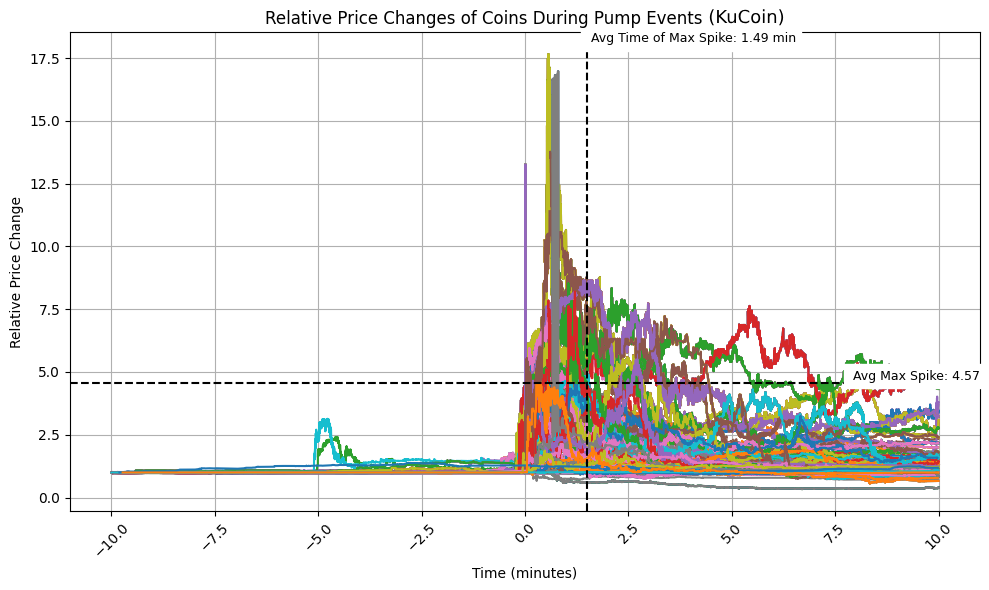}
        \caption{Relative Price Changes and Price Spike Metrics During KuCoin Pump Events.}
        \label{fig:relative_price_changes_kucoin}
    \end{minipage}%
    \hfill
    \begin{minipage}[t]{0.48\textwidth}
        \centering
        \includegraphics[width=\textwidth]{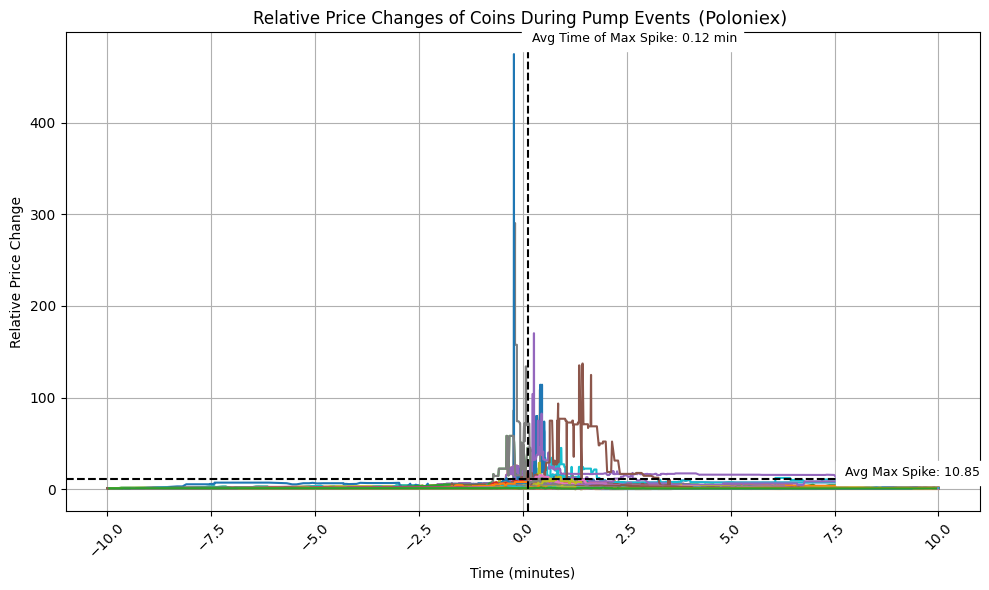}
        \caption{Relative Price Changes and Price Spike Metrics During Poloniex Pump Events.}
        \label{fig:relative_price_changes_poloniex}
    \end{minipage}
\end{figure*}

Figures~\ref{fig:relative_price_changes_kucoin} and~\ref{fig:relative_price_changes_poloniex} illustrate the relative, normalized price changes during pump events on KuCoin (approximately 700 events) and Poloniex (approximately 200 events), capturing a 20-minute window spanning ten minutes before to ten minutes after the start of each event.

On average, the time to reach the maximum price spike was calculated at 1.49 minutes for KuCoin and just 0.12 minutes for Poloniex. The magnitude of the price spikes, measured relative to a baseline (the midpoint between the bid and ask prices ten minutes prior to the pump), was notably higher on Poloniex (10.85) compared to KuCoin (4.57). These results suggest that pumps on Poloniex not only reach their peak more quickly but also result in larger relative price spikes.
This difference aligns with the observations from supplementary Figures~\ref{fig:avg_daily_volumes_by_channels_kucoin} and~\ref{fig:avg_daily_volumes_by_channels_poloniex} that show lower trading volumes and liquidity on Poloniex. Limited market depth and lower liquidity make it easier to induce extreme price fluctuations, leading to shorter-lived but sharper price spikes during pump events.


\subsubsection{Order Size}

Order size provides valuable insights into trader behavior during pump events, shedding light on patterns such as early positioning, cautious trading, and aggressive buying. Analyzing the average order sizes across different coins and exchanges reveals distinct behaviors and strategies.

On KuCoin, selected coins exhibit varying trends in order size during pump events. As shown in Table~\ref{tab:Kucoin}, KOL has the largest average order size before the pump, suggesting early trader positioning in anticipation of the event. Conversely, KPOL displays a significant increase in average order size during the pump, with an increase ratio of 1.30, indicating heightened interest and aggressive buying activity. IZI, pumped twice, shows a consistent decline in order size during the pump, likely due to profit-taking or cautious trading, as traders place smaller orders to mitigate risks in volatile conditions.

\begin{table}[htbp]
\centering
\caption{Average Order Sizes for Selected Coins on KuCoin During the 7 Days Before the Pump and on the Pump Day}
\label{tab:Kucoin}
\resizebox{\columnwidth}{!}{%
\begin{tabular}{l r r r}
\toprule
\textbf{Symbol} & \textbf{Avg Order Size (7 Days)} & \textbf{Avg Order Size (Pump Day)} & \textbf{Order Size Increase Ratio} \\
\midrule
AIEPK & 2257.07 & 2203.07 & 0.98 \\
IZI & 2199.62 & 752.08 & 0.35 \\
IZI & 2662.44 & 2130.59 & 0.80 \\
KOL & 4202.19 & 3471.90 & 0.83 \\
KPOL & 1750.89 & 2270.19 & 1.30 \\
MTS & 704.36 & 665.11 & 0.94 \\
\bottomrule
\end{tabular}
}
\end{table}

On Poloniex, market behavior during pump events is notably more volatile. Table~\ref{tab:Poloniex} highlights varied trends, with some coins displaying large order sizes before the pump, indicative of early positioning, while others experience sharp increases during the pump, reflecting sudden trader interest. For example, AMC shows inconsistent behavior, with stable order sizes in one instance and a dramatic increase in another, suggesting unpredictable trader dynamics. DMT and GFT demonstrate cautious trading, with declines in order size during the pump. Many coins on Poloniex experience periods of inactivity, signaling market hesitation or uncertainty leading up to the pump.

\begin{table}[htbp]
\centering
\caption{Average Order Sizes for Selected Coins on Poloniex During the 7 Days Before the Pump and on the Pump Day}
\label{tab:Poloniex}
\resizebox{\columnwidth}{!}{%
\begin{tabular}{l r r r}
\toprule
\textbf{Symbol} & \textbf{Avg Order Size (7 Days)} & \textbf{Avg Order Size (Pump Day)} & \textbf{Order Size Increase Ratio} \\
\midrule
AMC     & 51.69   & 49.05    & 0.95 \\
AMC     & 20.90   & 198.87   & 9.51 \\
COLLAB  & 914.13  & 884.18   & 0.97 \\
DMT     & 182.67  & 149.85   & 0.82 \\
GFT     & 37.24   & 529.65   & 14.03 \\
GFT     & 296.42  & 214.56   & 0.72 \\
GFT     & 808.65  & 775.09   & 0.96 \\
\bottomrule
\end{tabular}
}
\end{table}

Supplementary Figures~\ref{fig:kucoin_bar_chart} and~\ref{fig:poloniex_order_size} in Appendix visualize the differences in average order sizes across both exchanges. KuCoin exhibits more predictable patterns, with traders often positioning themselves ahead of pump events, as seen with KOL. In contrast, Poloniex shows greater variability, with some coins like COLLAB maintaining consistent order sizes, while others, such as AMC and GFT, demonstrate dramatic fluctuations during the pump.

\subsection{Filter Models to Reduce the Candidate Pool}
\label{sec:Filter}

Cryptocurrency exchanges often list thousands of tokens, creating significant scalability challenges for real-time tracking. For instance, MEXC lists 2,738 tokens \cite{mexc_global_coinranking}. To manage this complexity, our pipeline calculates and stores only order book and trade features in real-time, avoiding the need to store raw data. However, to further reduce the number of tokens tracked, we developed filter models to eliminate coins that lack the attributes commonly associated with pumped tokens.

Our filtering approach used a simple rule-based model to identify tokens within a specific market capitalization range. This range, spanning from 0 (unreported) to 60 million USD, targets lower market cap coins that are more vulnerable to price manipulation. As shown in Figure~\ref{fig:mcap_price}, the median market capitalization of pumped tokens is 2.7 million USD, with 95.7\% of the 1,431 historical pumped coins falling within this range. 

\section{Discussion}
\label{sec:Discussion}

The Z-score prediction model shows strong potential in predicting target coins for P\&D events, particularly when applied shortly before the pump. In our backtests, the target coin ranked within the top five by Z-score in 55.81\% of cases and within the top ten in 74.42\% when run 20 seconds prior to the event. However, performance declined significantly as the lead time increased, with TOP5 accuracy dropping to 41.45\% at 40 seconds and 19.05\% at 1 minute before the event. This suggests that critical trades or orders often occur just before the pump, making early detection more difficult.
Combining trade and order book data significantly improved the model’s performance, with a TOP5 accuracy of 55.81\% when using both, compared to 46.51\% with only trade data and 44.19\% with only order book data. This underscores the importance of leveraging complementary data sources for a more complete view of pre-pump market activity.

\subsubsection*{Limitations and Challenges}
The model's declining performance with increased lead time highlights its reliance on late-stage market signals. Detectable anomalies in trade and order book data typically emerge shortly before the target coin is announced, limiting the model’s ability to predict pumps well in advance. This reflects the orchestrators’ strategy of placing key orders just before publicizing the pump, creating a narrow window for detection.
Data sparsity also posed challenges, as many coins on Poloniex exhibit minimal or no activity. This sometimes reduced the comparison pool to fewer than 50 coins, potentially inflating the model’s performance by making it statistically easier for the target coin to rank higher. While this limitation may have biased results in cases with sparse data, the model's consistent performance in datasets with more active coins remains promising.

\subsubsection*{Potential Improvements and Future Research}
Future iterations could benefit from integrating on-chain transaction data, as P\&D organizers often transfer liquidity from other exchanges to the pump's target platform. Monitoring large or suspicious deposits into the target exchange could serve as an early warning signal.
If on-chain data is unavailable, expanding the analysis to include multiple exchanges could improve detection capabilities, albeit with increased computational demands. Additionally, incorporating broader market indicators, such as social media sentiment and external events, could enhance early detection accuracy.

Addressing data sparsity through a more robust comparison pool is another priority. Techniques such as jump analysis \cite{lee2008jumps} or fine-tuning time series transformer models \cite{rasul2024lagllamafoundationmodelsprobabilistic} could further refine anomaly detection. Leveraging pretrained time series models with labeled datasets offers another promising avenue for improvement.

A recent advancement in time series forecasting is the development of the Lag LLAMA model \cite{rasul2024lagllamafoundationmodelsprobabilistic}, which is based on the transformer architecture. This framework underlies sophisticated large language models such as ChatGPT and Claude. Transformer models excel at capturing long-range dependencies in sequential data, making them particularly suitable for a variety of predictive tasks.

Lag LLAMA has demonstrated promising results in predicting financial data \cite{bahelka2024comparativeanalysismixeddatasampling}. However, its current application is restricted to univariate data, which limits its utility for studies such as ours that involve multivariate datasets. Future extensions of the Lag LLAMA model to support multivariate data would open the door to new opportunities. Exploring its performance in these more complex, noisy, and interdependent settings could yield valuable insights and further refine predictive capabilities.


\subsubsection*{Susceptibility of Token Standards for P\&D Schemes}

\noindent
\textbf{Non-Fungible Token Standards.}
Non-fungible token (NFT) standards, such as ERC-721 tokens and Inscriptions \cite{Inscriptions}, are generally unsuitable for P\&D schemes. NFTs are unique and often linked to a specific digital or physical asset, making them incompatible with the mass trading needed for P\&D events. Unlike fungible tokens, where multiple parties can hold large quantities simultaneously, only one entity can own an NFT at any given time. 

Executing a P\&D scheme for NFTs would require orchestrators to resell the same token multiple times at progressively higher prices, which is logistically challenging and impractical. Such schemes would involve significant coordination, high gas fees, and blockchain congestion. Moreover, insider trading—common in P\&D schemes—would be nearly impossible, as orchestrators cannot acquire and hold unique tokens in advance for a mass sale.

Additionally, certain operational barriers exist for trading NFTs like Inscriptions, which require the "ord" client and a fully synchronized Bitcoin node \cite{nervos2024inscriptions}. Supporting this analysis, our historical dataset revealed no instances of NFTs being targeted in P\&D schemes. Telegram channels and marketplaces for NFTs, such as OpenSea and Magic Eden, lack any signs of coordinated P\&D activity.
Fragmented ownership of NFTs, which enables shared liquidity, might create conditions more conducive to P\&D schemes. While not observed in our data, other manipulation tactics, such as wash trading—where the same entity repeatedly buys and sells an NFT to inflate its price—are more likely in the NFT space.

\vspace{0.5em}
\noindent
\textbf{Fungible Token Standards.}
In contrast to NFTs, fungible tokens are inherently more susceptible to P\&D schemes due to their interchangeable nature. Standards such as ERC-20, BEP-20 \cite{BEP20}, and experimental types like BRC-20 \cite{BRC20} or Runes enable mass trading, making them ideal for orchestrators seeking to manipulate prices. These tokens allow multiple traders to hold large amounts simultaneously, facilitating coordinated buying during the pump phase and subsequent dumping for profit.
ERC-20-like tokens, in particular, are widely used due to their ease of listing on exchanges and large market volumes. Their relatively low entry barriers for traders and higher anonymity compared to NFTs make them prime targets for manipulation. Anonymity, afforded by high token supplies, conceals orchestrators’ actions and ensures smoother coordination.

All target coins in observed P\&D schemes in this study adhered to ERC-20-like standards. However, no pumps were observed involving listed BRC-20 tokens or Runes. This can be attributed to their limited trading availability on centralized exchanges \cite{Gate2024}, their experimental nature, and the lower adoption rates. Additionally, BRC-20 tokens face inefficiencies, including slower transaction speeds and higher fees, especially under network congestion \cite{li-2024}. These barriers reduce their attractiveness for orchestrators seeking rapid and large-scale price manipulation.

Nonetheless, the fungible nature of these newer standards implies potential susceptibility to P\&D schemes as their popularity and market adoption grow. With improved infrastructure and broader exchange listings, these tokens may become more prominent targets, warranting further research into their vulnerabilities.

\section{Conclusion}
\label{sec:Conclusion}

This study presents a robust framework for detecting pump-and-dump (P\&D) schemes in cryptocurrency markets, addressing a critical gap in research that predominantly focuses on post-event analysis or detecting pumps after they begin. By developing a real-time, ex-ante prediction pipeline, we enhance the ability to identify these schemes before they unfold.

Our approach integrates multiple data streams, including Telegram channel monitoring and high-frequency trade and order book data, to achieve accurate predictions. The Z-score-based model demonstrated strong predictive performance, correctly ranking the target coin within the top five in 55.81\% of cases and within the top ten in 74.42\% of cases when applied seconds before the pump event. While the model’s effectiveness diminishes as the prediction window extends, these results highlight the potential for early detection and its application in preemptive warning systems for centralized exchanges. We also provide a novel examination of token standards like BRC-20 and Runes, which we find not susceptible to the P\&D schemes. 

\bibliography{main}


\begin{thebibliography}{44}


\ifx \showCODEN    \undefined \def \showCODEN     #1{\unskip}     \fi
\ifx \showDOI      \undefined \def \showDOI       #1{#1}\fi
\ifx \showISBNx    \undefined \def \showISBNx     #1{\unskip}     \fi
\ifx \showISBNxiii \undefined \def \showISBNxiii  #1{\unskip}     \fi
\ifx \showISSN     \undefined \def \showISSN      #1{\unskip}     \fi
\ifx \showLCCN     \undefined \def \showLCCN      #1{\unskip}     \fi
\ifx \shownote     \undefined \def \shownote      #1{#1}          \fi
\ifx \showarticletitle \undefined \def \showarticletitle #1{#1}   \fi
\ifx \showURL      \undefined \def \showURL       {\relax}        \fi
\providecommand\bibfield[2]{#2}
\providecommand\bibinfo[2]{#2}
\providecommand\natexlab[1]{#1}
\providecommand\showeprint[2][]{arXiv:#2}

\bibitem[Coi(2024)]%
        {CoinMarketCap}
 \bibinfo{year}{2024}\natexlab{}.
\newblock \bibinfo{title}{CoinMarketCap}.
\newblock \bibinfo{howpublished}{\url{https://coinmarketcap.com/}}.
\newblock
\newblock
\shownote{Accessed: 2024-09-10}.


\bibitem[AlphaPoint(2024)]%
        {alphapoint2024erc20vsbep20}
\bibfield{author}{\bibinfo{person}{AlphaPoint}.} \bibinfo{year}{2024}\natexlab{}.
\newblock \bibinfo{title}{ERC20 vs. BEP-20: What’s the Difference?}
\newblock
\newblock
\urldef\tempurl%
\url{https://alphapoint.com/blog/erc20-vs-bep20-differences/}
\showURL{%
\tempurl}
\newblock
\shownote{[Online; accessed 04-October-2024]}.


\bibitem[Bahelka and de~Weerd(2024)]%
        {bahelka2024comparativeanalysismixeddatasampling}
\bibfield{author}{\bibinfo{person}{Adam Bahelka} {and} \bibinfo{person}{Harmen de Weerd}.} \bibinfo{year}{2024}\natexlab{}.
\newblock \bibinfo{title}{Comparative analysis of Mixed-Data Sampling (MIDAS) model compared to Lag-Llama model for inflation nowcasting}.
\newblock
\newblock
\showeprint[arxiv]{2407.08510}~[econ.EM]
\urldef\tempurl%
\url{https://arxiv.org/abs/2407.08510}
\showURL{%
\tempurl}


\bibitem[Barrington and Merrill(2022)]%
        {barrington2022erc-721}
\bibfield{author}{\bibinfo{person}{Sarah Barrington} {and} \bibinfo{person}{Nick Merrill}.} \bibinfo{year}{2022}\natexlab{}.
\newblock \bibinfo{title}{The Fungibility of Non-Fungible Tokens: A Quantitative Analysis of ERC-721 Metadata}.
\newblock
\newblock
\showeprint[arxiv]{2209.14517}~[cs.CR]
\urldef\tempurl%
\url{https://arxiv.org/abs/2209.14517}
\showURL{%
\tempurl}


\bibitem[Binance(nd)]%
        {BEP20}
\bibfield{author}{\bibinfo{person}{Binance}.} \bibinfo{year}{n.d.}\natexlab{}.
\newblock \bibinfo{title}{{BEP-20: Binance Smart Chain Token Standard}}.
\newblock \bibinfo{howpublished}{\url{https://docs.binance.org/smart-chain/developer/BEP20.html}}.
\newblock
\newblock
\shownote{Accessed: Oct. 22, 2024}.


\bibitem[Chainalysis(2023)]%
        {chainalysis2023crypto}
\bibfield{author}{\bibinfo{person}{Chainalysis}.} \bibinfo{year}{2023}\natexlab{}.
\newblock \bibinfo{title}{Crypto Pump and Dump Schemes Make Up 24\% of New Tokens}.
\newblock
\newblock
\urldef\tempurl%
\url{https://www.chainalysis.com/blog/2022-crypto-pump-and-dump-schemes/}
\showURL{%
\tempurl}
\newblock
\shownote{Accessed: 2024-05-19}.


\bibitem[{Chang, Yupeng and Wang, Xu and Wang, Jindong and Wu, Yuan and Yang, Linyi and Zhu, Kaijie and Chen, Hao and Yi, Xiaoyuan and Wang, Cunxiang and Wang, Yidong and Ye, Wei and Zhang, Yue and Chang, Yi and Yu, Philip S. and Yang, Qiang and Xie, Xing}(2024)]%
        {2024SurveyLLM}
\bibfield{author}{\bibinfo{person}{{Chang, Yupeng and Wang, Xu and Wang, Jindong and Wu, Yuan and Yang, Linyi and Zhu, Kaijie and Chen, Hao and Yi, Xiaoyuan and Wang, Cunxiang and Wang, Yidong and Ye, Wei and Zhang, Yue and Chang, Yi and Yu, Philip S. and Yang, Qiang and Xie, Xing}}.} \bibinfo{year}{2024}\natexlab{}.
\newblock \showarticletitle{A Survey on Evaluation of Large Language Models}.
\newblock \bibinfo{journal}{\emph{ACM Trans. Intell. Syst. Technol.}} \bibinfo{volume}{15}, \bibinfo{number}{3}, Article \bibinfo{articleno}{39} (\bibinfo{date}{March} \bibinfo{year}{2024}), \bibinfo{numpages}{45}~pages.
\newblock
\showISSN{2157-6904}
\urldef\tempurl%
\url{https://doi.org/10.1145/3641289}
\showDOI{\tempurl}


\bibitem[CoinCodex(2024)]%
        {CoinCodex}
\bibfield{author}{\bibinfo{person}{CoinCodex}.} \bibinfo{year}{2024}\natexlab{}.
\newblock \bibinfo{title}{CoinCodex}.
\newblock \bibinfo{howpublished}{\url{https://coincodex.com/}}.
\newblock
\newblock
\shownote{Accessed: 2024-09-10}.


\bibitem[Coinranking(2024)]%
        {mexc_global_coinranking}
\bibfield{author}{\bibinfo{person}{Coinranking}.} \bibinfo{year}{2024}\natexlab{}.
\newblock \bibinfo{title}{MEXC Global - Coins}.
\newblock
\newblock
\urldef\tempurl%
\url{https://coinranking.com/exchange/YLfc3FfDe+mexc-global/coins}
\showURL{%
\tempurl}
\newblock
\shownote{Accessed: 2024-09-09}.


\bibitem[Domo(2023)]%
        {BRC20}
\bibfield{author}{\bibinfo{person}{Domo}.} \bibinfo{year}{2023}\natexlab{}.
\newblock \bibinfo{title}{{BRC-20: An Experimental Fungible Token Standard Using Ordinals}}.
\newblock \bibinfo{howpublished}{\url{https://domo-2.gitbook.io/brc-20-experiment/}}.
\newblock
\newblock
\shownote{Accessed: Oct. 22, 2024}.


\bibitem[Entriken et~al\mbox{.}(2018)]%
        {ERC721}
\bibfield{author}{\bibinfo{person}{William Entriken}, \bibinfo{person}{Dieter Shirley}, \bibinfo{person}{Jacob Evans}, {and} \bibinfo{person}{Nastassia Sachs}.} \bibinfo{year}{2018}\natexlab{}.
\newblock \bibinfo{title}{{ERC-721: Non-Fungible Token Standard}}.
\newblock \bibinfo{howpublished}{\textit{Ethereum Improvement Proposals}, no. 721}.
\newblock
\newblock
\shownote{\url{https://eips.ethereum.org/EIPS/eip-721}}.


\bibitem[Ester et~al\mbox{.}(1996)]%
        {ester1996density}
\bibfield{author}{\bibinfo{person}{Martin Ester}, \bibinfo{person}{Hans-Peter Kriegel}, \bibinfo{person}{Jörg Sander}, {and} \bibinfo{person}{Xiaowei Xu}.} \bibinfo{year}{1996}\natexlab{}.
\newblock \showarticletitle{A density-based algorithm for discovering clusters in large spatial databases with noise}. In \bibinfo{booktitle}{\emph{Proceedings of the 2nd International Conference on Knowledge Discovery and Data Mining (KDD)}}. \bibinfo{pages}{226--231}.
\newblock


\bibitem[Freni et~al\mbox{.}(2020)]%
        {9219709}
\bibfield{author}{\bibinfo{person}{Pierluigi Freni}, \bibinfo{person}{Enrico Ferro}, {and} \bibinfo{person}{Roberto Moncada}.} \bibinfo{year}{2020}\natexlab{}.
\newblock \showarticletitle{Tokenization and Blockchain Tokens Classification: a morphological framework}. In \bibinfo{booktitle}{\emph{2020 IEEE Symposium on Computers and Communications (ISCC)}}. \bibinfo{pages}{1--6}.
\newblock
\urldef\tempurl%
\url{https://doi.org/10.1109/ISCC50000.2020.9219709}
\showDOI{\tempurl}


\bibitem[Gogol et~al\mbox{.}(2024)]%
        {gogol2024sokdefi}
\bibfield{author}{\bibinfo{person}{Krzysztof Gogol}, \bibinfo{person}{Christian Killer}, \bibinfo{person}{Malte Schlosser}, \bibinfo{person}{Thomas Bocek}, \bibinfo{person}{Burkhard Stiller}, {and} \bibinfo{person}{Claudio Tessone}.} \bibinfo{year}{2024}\natexlab{}.
\newblock \showarticletitle{{SoK: Decentralized Finance (DeFi) - Fundamentals, Taxonomy and Risks}}.
\newblock \bibinfo{journal}{\emph{arXiv 2404.11281}} (\bibinfo{year}{2024}).
\newblock
\urldef\tempurl%
\url{https://arxiv.org/abs/2404.11281}
\showURL{%
\tempurl}


\bibitem[Hu et~al\mbox{.}(2023)]%
        {hu2023sequence}
\bibfield{author}{\bibinfo{person}{Sihao Hu}, \bibinfo{person}{Zhen Zhang}, \bibinfo{person}{Shengliang Lu}, \bibinfo{person}{Bingsheng He}, {and} \bibinfo{person}{Zhao Li}.} \bibinfo{year}{2023}\natexlab{}.
\newblock \showarticletitle{Sequence-based target coin prediction for cryptocurrency pump-and-dump}.
\newblock \bibinfo{journal}{\emph{Proceedings of the ACM on Management of Data}} \bibinfo{volume}{1}, \bibinfo{number}{1} (\bibinfo{year}{2023}), \bibinfo{pages}{1--19}.
\newblock


\bibitem[Inc.(2024)]%
        {lunarcrush2024}
\bibfield{author}{\bibinfo{person}{LunarCrush Inc.}} \bibinfo{year}{2024}\natexlab{}.
\newblock \bibinfo{title}{LunarCrush}.
\newblock \bibinfo{howpublished}{\url{https://lunarcrush.com/home}}.
\newblock
\newblock
\shownote{[Online; accessed 10-October-2024]}.


\bibitem[Kalyan(2024)]%
        {2024SurveyGTP3}
\bibfield{author}{\bibinfo{person}{Katikapalli~Subramanyam Kalyan}.} \bibinfo{year}{2024}\natexlab{}.
\newblock \showarticletitle{{A survey of GPT-3 family large language models including ChatGPT and GPT-4}}.
\newblock \bibinfo{journal}{\emph{Natural Language Processing Journal}}  \bibinfo{volume}{6} (\bibinfo{year}{2024}), \bibinfo{pages}{100048}.
\newblock
\showISSN{2949-7191}
\urldef\tempurl%
\url{https://doi.org/10.1016/j.nlp.2023.100048}
\showDOI{\tempurl}


\bibitem[Kamps and Kleinberg(2018)]%
        {kamps2018moon}
\bibfield{author}{\bibinfo{person}{Josh Kamps} {and} \bibinfo{person}{Bennett Kleinberg}.} \bibinfo{year}{2018}\natexlab{}.
\newblock \showarticletitle{To the moon: defining and detecting cryptocurrency pump-and-dumps}.
\newblock \bibinfo{journal}{\emph{Crime Science}} \bibinfo{volume}{7}, \bibinfo{number}{1} (\bibinfo{year}{2018}), \bibinfo{pages}{1--18}.
\newblock


\bibitem[Kroitor and contributors(nd)]%
        {ccxt}
\bibfield{author}{\bibinfo{person}{Igor Kroitor} {and} \bibinfo{person}{contributors}.} \bibinfo{year}{n.d.}\natexlab{}.
\newblock \bibinfo{title}{{CCXT: Cryptocurrency Trading Library}}.
\newblock \bibinfo{howpublished}{\url{https://github.com/ccxt/ccxt}}.
\newblock


\bibitem[{KuCoin}(2024)]%
        {kucoin2024runes}
\bibfield{author}{\bibinfo{person}{{KuCoin}}.} \bibinfo{year}{2024}\natexlab{}.
\newblock \bibinfo{title}{What Is Runes Protocol? Bitcoin’s Latest Fungible Token Standard}.
\newblock
\newblock
\urldef\tempurl%
\url{https://www.kucoin.com/learn/crypto/what-is-runes-protocol-on-bitcoin}
\showURL{%
\tempurl}
\newblock
\shownote{Accessed: 2024-10-05}.


\bibitem[La~Morgia et~al\mbox{.}(2023)]%
        {la2023doge}
\bibfield{author}{\bibinfo{person}{Massimo La~Morgia}, \bibinfo{person}{Alessandro Mei}, \bibinfo{person}{Francesco Sassi}, {and} \bibinfo{person}{Julinda Stefa}.} \bibinfo{year}{2023}\natexlab{}.
\newblock \showarticletitle{The doge of wall street: Analysis and detection of pump and dump cryptocurrency manipulations}.
\newblock \bibinfo{journal}{\emph{ACM Transactions on Internet Technology}} \bibinfo{volume}{23}, \bibinfo{number}{1} (\bibinfo{year}{2023}), \bibinfo{pages}{1--28}.
\newblock


\bibitem[Lee and Mykland(2008)]%
        {lee2008jumps}
\bibfield{author}{\bibinfo{person}{Suzanne~S Lee} {and} \bibinfo{person}{Per~A Mykland}.} \bibinfo{year}{2008}\natexlab{}.
\newblock \showarticletitle{Jumps in financial markets: A new nonparametric test and jump dynamics}.
\newblock \bibinfo{journal}{\emph{The Review of Financial Studies}} \bibinfo{volume}{21}, \bibinfo{number}{6} (\bibinfo{year}{2008}), \bibinfo{pages}{2535--2563}.
\newblock


\bibitem[Li et~al\mbox{.}(2024)]%
        {li-2024}
\bibfield{author}{\bibinfo{person}{Ningran Li}, \bibinfo{person}{Minfeng Qi}, \bibinfo{person}{Qin Wang}, {and} \bibinfo{person}{Shiping Chen}.} \bibinfo{year}{2024}\natexlab{}.
\newblock \showarticletitle{{Bitcoin Inscriptions: Foundations and Beyond}}.
\newblock \bibinfo{journal}{\emph{arXiv (Cornell University)}} (\bibinfo{date}{1} \bibinfo{year}{2024}).
\newblock
\urldef\tempurl%
\url{https://doi.org/10.48550/arxiv.2401.17581}
\showDOI{\tempurl}


\bibitem[MacQueen(1967)]%
        {macqueen1967kmeans}
\bibfield{author}{\bibinfo{person}{J. MacQueen}.} \bibinfo{year}{1967}\natexlab{}.
\newblock \showarticletitle{Some methods for classification and analysis of multivariate observations}. In \bibinfo{booktitle}{\emph{Proceedings of the Fifth Berkeley Symposium on Mathematical Statistics and Probability}}. \bibinfo{pages}{281--297}.
\newblock


\bibitem[Mansourifar et~al\mbox{.}(2020)]%
        {mansourifar2020hybrid}
\bibfield{author}{\bibinfo{person}{Hadi Mansourifar}, \bibinfo{person}{Lin Chen}, {and} \bibinfo{person}{Weidong Shi}.} \bibinfo{year}{2020}\natexlab{}.
\newblock \showarticletitle{Hybrid cryptocurrency pump and dump detection}.
\newblock \bibinfo{journal}{\emph{arXiv preprint arXiv:2003.06551}} (\bibinfo{year}{2020}).
\newblock


\bibitem[Messias et~al\mbox{.}(2024)]%
        {messias2024Inscription}
\bibfield{author}{\bibinfo{person}{Johnnatan Messias}, \bibinfo{person}{Krzysztof Gogol}, \bibinfo{person}{Maria~Inês Silva}, {and} \bibinfo{person}{Benjamin Livshits}.} \bibinfo{year}{2024}\natexlab{}.
\newblock \showarticletitle{The Writing is on the Wall: Analyzing the Boom of Inscriptions and its Impact on EVM-compatible Blockchains}.
\newblock \bibinfo{journal}{\emph{arXiv 2405.15288}} (\bibinfo{year}{2024}).
\newblock
\urldef\tempurl%
\url{https://arxiv.org/abs/2405.15288}
\showURL{%
\tempurl}


\bibitem[Morales et~al\mbox{.}(2023)]%
        {morales2023erc-20}
\bibfield{author}{\bibinfo{person}{Alfredo~J. Morales}, \bibinfo{person}{Shahar Somin}, \bibinfo{person}{Yaniv Altshuler}, {and} \bibinfo{person}{Alex~Sandy Pentland}.} \bibinfo{year}{2023}\natexlab{}.
\newblock \showarticletitle{Patterns of User Behavior and Token Adoption on ERC20}.
\newblock \bibinfo{journal}{\emph{SN Computer Science}} \bibinfo{volume}{4}, \bibinfo{number}{6} (\bibinfo{date}{Sept.} \bibinfo{year}{2023}), \bibinfo{pages}{753}.
\newblock
\urldef\tempurl%
\url{https://doi.org/10.1007/s42979-023-02200-6}
\showURL{%
\tempurl}


\bibitem[Murtagh and Contreras(2012)]%
        {murtagh2012agglomerative}
\bibfield{author}{\bibinfo{person}{F. Murtagh} {and} \bibinfo{person}{P. Contreras}.} \bibinfo{year}{2012}\natexlab{}.
\newblock \showarticletitle{Algorithms for hierarchical clustering: an overview}.
\newblock \bibinfo{journal}{\emph{Wiley Interdisciplinary Reviews: Data Mining and Knowledge Discovery}} \bibinfo{volume}{2}, \bibinfo{number}{1} (\bibinfo{year}{2012}), \bibinfo{pages}{86--97}.
\newblock
\urldef\tempurl%
\url{https://doi.org/10.1002/widm.53}
\showDOI{\tempurl}


\bibitem[{Nervos Network}(2024)]%
        {nervos2024inscriptions}
\bibfield{author}{\bibinfo{person}{{Nervos Network}}.} \bibinfo{year}{2024}\natexlab{}.
\newblock \bibinfo{title}{Guide to Inscriptions}.
\newblock \bibinfo{howpublished}{\url{https://www.nervos.org/knowledge-base/guide_to_inscriptions}}.
\newblock
\newblock
\shownote{Accessed: 2024-08-28}.


\bibitem[Nguyen et~al\mbox{.}(2020)]%
        {nguyen2020bertweetpretrainedlanguagemodel}
\bibfield{author}{\bibinfo{person}{Dat~Quoc Nguyen}, \bibinfo{person}{Thanh Vu}, {and} \bibinfo{person}{Anh~Tuan Nguyen}.} \bibinfo{year}{2020}\natexlab{}.
\newblock \bibinfo{title}{BERTweet: A pre-trained language model for English Tweets}.
\newblock
\newblock
\showeprint[arxiv]{2005.10200}~[cs.CL]
\urldef\tempurl%
\url{https://arxiv.org/abs/2005.10200}
\showURL{%
\tempurl}


\bibitem[Pilkington(2016)]%
        {pilkington2016blockchain}
\bibfield{author}{\bibinfo{person}{Marc Pilkington}.} \bibinfo{year}{2016}\natexlab{}.
\newblock \showarticletitle{Blockchain Technology: Principles and Applications}.
\newblock In \bibinfo{booktitle}{\emph{Research Handbook on Digital Transformations}}, \bibfield{editor}{\bibinfo{person}{F.~Xavier Olleros} {and} \bibinfo{person}{Majlinda Zhegu}} (Eds.). \bibinfo{publisher}{Edward Elgar}, \bibinfo{pages}{225--253}.
\newblock
\urldef\tempurl%
\url{https://ssrn.com/abstract=2662660}
\showURL{%
\tempurl}


\bibitem[Rasul et~al\mbox{.}(2024)]%
        {rasul2024lagllamafoundationmodelsprobabilistic}
\bibfield{author}{\bibinfo{person}{Kashif Rasul}, \bibinfo{person}{Arjun Ashok}, \bibinfo{person}{Andrew~Robert Williams}, \bibinfo{person}{Hena Ghonia}, \bibinfo{person}{Rishika Bhagwatkar}, \bibinfo{person}{Arian Khorasani}, \bibinfo{person}{Mohammad Javad~Darvishi Bayazi}, \bibinfo{person}{George Adamopoulos}, \bibinfo{person}{Roland Riachi}, \bibinfo{person}{Nadhir Hassen}, \bibinfo{person}{Marin Biloš}, \bibinfo{person}{Sahil Garg}, \bibinfo{person}{Anderson Schneider}, \bibinfo{person}{Nicolas Chapados}, \bibinfo{person}{Alexandre Drouin}, \bibinfo{person}{Valentina Zantedeschi}, \bibinfo{person}{Yuriy Nevmyvaka}, {and} \bibinfo{person}{Irina Rish}.} \bibinfo{year}{2024}\natexlab{}.
\newblock \bibinfo{title}{Lag-Llama: Towards Foundation Models for Probabilistic Time Series Forecasting}.
\newblock
\newblock
\showeprint[arxiv]{2310.08278}~[cs.LG]
\urldef\tempurl%
\url{https://arxiv.org/abs/2310.08278}
\showURL{%
\tempurl}


\bibitem[Reynolds(2009)]%
        {reynolds2009gmm}
\bibfield{author}{\bibinfo{person}{D.~A. Reynolds}.} \bibinfo{year}{2009}\natexlab{}.
\newblock \showarticletitle{Gaussian Mixture Models}.
\newblock In \bibinfo{booktitle}{\emph{Encyclopedia of Biometrics}}. \bibinfo{publisher}{Springer}, \bibinfo{pages}{827--832}.
\newblock
\urldef\tempurl%
\url{https://doi.org/10.1007/978-0-387-73003-5_196}
\showDOI{\tempurl}


\bibitem[Rodarmor(2023a)]%
        {Inscriptions}
\bibfield{author}{\bibinfo{person}{Casey Rodarmor}.} \bibinfo{year}{2023}\natexlab{a}.
\newblock \bibinfo{title}{{Inscribing Mainnet}}.
\newblock \bibinfo{howpublished}{\url{https://rodarmor.com/blog/inscribing-mainnet/}}.
\newblock
\newblock
\shownote{Accessed: Oct. 22, 2024}.


\bibitem[Rodarmor(2023b)]%
        {rodarmor-2023}
\bibfield{author}{\bibinfo{person}{Casey Rodarmor}.} \bibinfo{year}{2023}\natexlab{b}.
\newblock \bibinfo{title}{The Ordinals Protocol}.
\newblock
\newblock
\urldef\tempurl%
\url{https://docs.ordinals.com}
\showURL{%
\tempurl}
\newblock
\shownote{Accessed: 2024-08-14}.


\bibitem[Rodarmor(2023c)]%
        {rodarmor2023runes}
\bibfield{author}{\bibinfo{person}{Casey Rodarmor}.} \bibinfo{year}{2023}\natexlab{c}.
\newblock \bibinfo{title}{Runes}.
\newblock
\newblock
\urldef\tempurl%
\url{https://rodarmor.com/blog/runes/}
\showURL{%
\tempurl}
\newblock
\shownote{Accessed: 2024-10-05}.


\bibitem[Team(2024)]%
        {Gate2024}
\bibfield{author}{\bibinfo{person}{Gate.io Team}.} \bibinfo{year}{2024}\natexlab{}.
\newblock \bibinfo{title}{Gate.io will list Runes Project \$DOG$\cdot$GO$\cdot$TO$\cdot$THE$\cdot$MOON(DOG)}.
\newblock
\newblock
\urldef\tempurl%
\url{https://www.gate.io/announcements/article/36191}
\showURL{%
\tempurl}
\newblock
\shownote{Accessed: 2024-10-17}.


\bibitem[{TGStat}(2024)]%
        {tgstat}
\bibfield{author}{\bibinfo{person}{{TGStat}}.} \bibinfo{year}{2024}\natexlab{}.
\newblock \bibinfo{title}{TGStat: Telegram Analytics}.
\newblock \bibinfo{howpublished}{\url{https://tgstat.com/}}.
\newblock
\newblock
\shownote{Accessed: 2024-09-09}.


\bibitem[Vermaak(2023)]%
        {vermaak2023runes}
\bibfield{author}{\bibinfo{person}{Werner Vermaak}.} \bibinfo{year}{2023}\natexlab{}.
\newblock \bibinfo{title}{What Is the Runes Protocol?}
\newblock
\newblock
\urldef\tempurl%
\url{https://coinmarketcap.com/academy/article/what-is-the-runes-protocol}
\showURL{%
\tempurl}
\newblock
\shownote{Accessed: 2024-10-05}.


\bibitem[Victor and Hagemann(2019)]%
        {victor2019cryptocurrency}
\bibfield{author}{\bibinfo{person}{Friedhelm Victor} {and} \bibinfo{person}{Tanja Hagemann}.} \bibinfo{year}{2019}\natexlab{}.
\newblock \showarticletitle{Cryptocurrency pump and dump schemes: Quantification and detection}. In \bibinfo{booktitle}{\emph{2019 International Conference on Data Mining Workshops (ICDMW)}}. IEEE, \bibinfo{pages}{244--251}.
\newblock


\bibitem[Vogelsteller and Buterin(2015)]%
        {ERC20}
\bibfield{author}{\bibinfo{person}{Fabian Vogelsteller} {and} \bibinfo{person}{Vitalik Buterin}.} \bibinfo{year}{2015}\natexlab{}.
\newblock \bibinfo{title}{{ERC-20: Token Standard}}.
\newblock \bibinfo{howpublished}{\textit{Ethereum Improvement Proposals}, no. 20}.
\newblock
\newblock
\shownote{\url{https://eips.ethereum.org/EIPS/eip-20}}.


\bibitem[Wang et~al\mbox{.}(2024)]%
        {wang2024survey}
\bibfield{author}{\bibinfo{person}{Lei Wang}, \bibinfo{person}{Chen Ma}, \bibinfo{person}{Xueyang Feng}, \bibinfo{person}{Zeyu Zhang}, \bibinfo{person}{Hao Yang}, \bibinfo{person}{Jingsen Zhang}, \bibinfo{person}{Zhiyuan Chen}, \bibinfo{person}{Jiakai Tang}, \bibinfo{person}{Xu Chen}, \bibinfo{person}{Yankai Lin}, {et~al\mbox{.}}} \bibinfo{year}{2024}\natexlab{}.
\newblock \showarticletitle{{A survey on large language model based autonomous agents}}.
\newblock \bibinfo{journal}{\emph{Frontiers of Computer Science}} \bibinfo{volume}{18}, \bibinfo{number}{6} (\bibinfo{year}{2024}), \bibinfo{pages}{186345}.
\newblock


\bibitem[Wołk(2019)]%
        {wolk2019}
\bibfield{author}{\bibinfo{person}{Krzysztof Wołk}.} \bibinfo{year}{2019}\natexlab{}.
\newblock \showarticletitle{Advanced social media sentiment analysis for short‐term cryptocurrency price prediction}.
\newblock \bibinfo{journal}{\emph{Expert Systems}}  \bibinfo{volume}{37} (\bibinfo{date}{11} \bibinfo{year}{2019}).
\newblock
\urldef\tempurl%
\url{https://doi.org/10.1111/exsy.12493}
\showDOI{\tempurl}


\bibitem[Xu and Livshits(2019)]%
        {xu2019anatomy}
\bibfield{author}{\bibinfo{person}{Jiahua Xu} {and} \bibinfo{person}{Benjamin Livshits}.} \bibinfo{year}{2019}\natexlab{}.
\newblock \showarticletitle{The anatomy of a cryptocurrency $\{$Pump-and-Dump$\}$ scheme}. In \bibinfo{booktitle}{\emph{28th USENIX Security Symposium (USENIX Security 19)}}. \bibinfo{pages}{1609--1625}.
\newblock


\end{thebibliography}

\appendix

\section{Appendix: Natural Language Processing}
\subsection{OpenAI Settings and Prompts}
\label{app:Prompts}
For labeling and information extraction, we utilized the latest version of GPT-4o. To improve consistency and reliability, the model's randomness was reduced by setting the temperature parameter to 0.2 and the top\_p parameter to 0.95. All other parameters remained at their default values.

\subsubsection{Labeling Prompt}
To classify Telegram messages from pump-and-dump channels, the model was instructed to assign one of six predefined labels. Each label was accompanied by a concise description, and the output format was specified as JSON. The following prompt was used:

\begin{lstlisting}
TASK: Label Telegram messages from pump-and-dump channels using the following labels:

0: Formal Announcement - Includes date, time, and exchange name.  
1: Pump Countdown - Countdown details, possibly with date/time.  
2: Target Coin - Specifies the token symbol to buy.  
3: Post Analysis - Reports profits or past pump summaries.  
4: Pump Canceled - Announces cancellation, delay, or postponement.  
5: Rest - Messages unrelated to the above, e.g., advertisements.

JSON FORMAT: {return_format}

MESSAGE: {message}
\end{lstlisting}

\subsubsection{Extraction Prompts}
The model also processed labeled messages to extract structured information. Two categories were targeted: announcement messages (extracting exchange, trading pair, and pump timing) and postponement/cancellation messages (updating the pump status). 

\paragraph{Announcement Extraction Prompt}
This prompt guided the model to extract essential details from pump announcements, as shown below:

\begin{lstlisting}
TASK: Extract exchange name, trading pair, and pump timing from the following message.  

INSTRUCTIONS:
1. Exchange: Identify the exchange name (marked by _CEX or _DEX).  
2. Trading Pair: Extract the base currency symbol; if not found, return null.  
3. Date & Time: Extract pump timing in ISO 8601 format (YYYY-MM-DD HH:MM\pmHH:MM).  
   - Use the message timestamp for inference if necessary, setting "inferred" to 1.  
   - If extraction or inference fails, return null.  

JSON FORMAT: {response_format}
\end{lstlisting}

\paragraph{Postponement/Cancellation Extraction Prompt}
This prompt instructed the model to determine if a pump was postponed or canceled and to extract updated information as needed:

\begin{lstlisting}
TASK: Determine if a pump is canceled or postponed and extract relevant information.

INSTRUCTIONS:
1. Cancellation/Delay:
   - If canceled, set "cancelled" to 1, otherwise 0.
   - If postponed, set "postponed" to 1 and extract the new pump timing.

2. New Pump Timing:
   - Calculate the new date/time based on delay information.
   - Return in ISO 8601 format (YYYY-MM-DD HH:MM\pmHH:MM).
   - If extraction or inference fails, return null.

JSON FORMAT: {response_format}
\end{lstlisting}


\section{Computed Metrics}
\subsection{Order Book and Trade Data Metrics}

Table \ref{tab:metrics} defines the key metrics derived from order book and trade data, including their formulas and a brief explanation.

\begin{table}[h]
\centering
\caption{Computed Metrics for Orderbook and Trade Data}
\label{tab:metrics}
\begin{tabular}{p{3cm}|p{5cm}}
\hline
\textbf{Metric} & \textbf{Description and Formula} \\
\hline
Bid-Ask Spread & 
Difference between the lowest ask price and highest bid price: 
\( P_{\text{min}}^{\text{ask}} - P_{\text{max}}^{\text{bid}} \) \\
\hline
Average Order Size & 
Mean size per order: \( \frac{1}{n} \sum_{i=1}^{n} Q_i \) \\
\hline
Imbalance & 
Ratio of total bid quantity to total ask quantity: 
\( \frac{\sum Q_{\text{bid}}}{\sum Q_{\text{ask}}} \) \\
\hline
Imbalance Ratio & 
Adjusted imbalance around mid-price \( P_{\text{mid}} \): 
\( \frac{(P_{\text{bid}} \cdot Q_{\text{bid}} \cdot P_{\text{mid}}) - (P_{\text{ask}} \cdot Q_{\text{ask}} \cdot P_{\text{mid}})}{P_{\text{bid}} \cdot Q_{\text{bid}} + P_{\text{ask}} \cdot Q_{\text{ask}}} \), 
where \( P_{\text{mid}} = \frac{P_{\text{max}} + P_{\text{min}}}{2} \) \\
\hline
Order Book Pressure & 
Proportion of bid quantity to total quantity: 
\( \frac{\sum Q_{\text{bid}}}{\sum Q_{\text{bid}} + \sum Q_{\text{ask}}} \) \\
\hline
Order Book Slope & 
Median of bid-ask quantity differences at price levels: 
\( \text{median} (\Delta Q_{\text{bid}} - \Delta Q_{\text{ask}}) \) \\
\hline
Liquidity Consumption & 
Total executed order quantity: \( \sum Q_{\text{executed}} \) \\
\hline
Order Flow Imbalance (OFI) & 
Difference between total bid and ask quantities: 
\( \sum Q_{\text{bid}} - \sum Q_{\text{ask}} \) \\
\hline
Market Orders Impact & 
Sum of market order quantities on bid and ask sides: 
\( \sum Q_{\text{bid, market}} + \sum Q_{\text{ask, market}} \) \\
\hline
Relative Impact & 
Relative price change at liquidity levels: 
\( \frac{P_{\text{after}} - P_{\text{before}}}{P_{\text{before}}} \), 
where \( P_{\text{before}} \) and \( P_{\text{after}} \) are prices before and after consumption. \\
\hline
\end{tabular}
\end{table}

\section{Appendix: Data and Model Evaluation}

\subsection{XGBoost Model}
\label{app:XGBoost}
This section contains the features we found to be most effective for the classification of coins into \textit{pumpable} and \textit{non-pumpable}. Initially, we calculated 27 features from the OHLCV (open, high, low, close, volume), market capitalization, and social time series data, which we collected for a sample of 1,714 coins (643 pumped coins and a random sample of 1,071 non-pump coins). The time series were made stationary by calculating the difference between consecutive terms in the series:

\begin{equation}
y_t' = y_t - y_{t-1}
\end{equation}

Additionally, each time series was normalized by dividing it by Bitcoin's corresponding value to remove differences over time. We chose Bitcoin because of its property as a central asset in the crypto market. Lastly, we applied min-max scaling across coins to normalize the scale of each feature while preserving inter-coin differences.

\subsubsection{Feature Selection}
From the initial 27 time series features, we calculated multiple features per time series using the Python library TS Fresh, as XGBoost cannot work with data in time series format. We applied recursive feature elimination with a 5-fold cross-validation, using the weighted F1 score as the evaluation metric. Table \ref{tab:key_features} lists the best features selected for each XGBoost model described in Section \ref{sec:Filter}.

\begin{table}[htbp]
\centering
\caption{Features and Descriptions for XGBoost Models 0 to 3}
\label{tab:key_features}
\resizebox{\columnwidth}{!}{%
\begin{tabular}{@{}p{3cm} p{5cm} c c c c@{}}
\toprule
\textbf{Feature} & \textbf{Description} & \textbf{Model 0} & \textbf{Model 1} & \textbf{Model 2} & \textbf{Model 3} \\ \midrule
Close Price (Max) & Highest recorded closing price & X & X & X & X \\ \midrule
Close Price (Min) & Lowest recorded closing price & X & X & X & X \\ \midrule
Market Cap (Median) & Median market capitalization & X & X & X & X \\ \midrule
Market Cap (Std. Dev.) & Volatility of market capitalization & X & X & X & X \\ \midrule
Market Cap (Max) & Maximum market capitalization & X & X & X & X \\ \midrule
Market Cap (Min) & Minimum market capitalization & X & X & X & X \\ \midrule
Trading Vol. (Max) & Maximum trading volume & X &  &  &  \\ \midrule
Close Price (Sum) & Total closing price sum &  & X &  &  \\ \midrule
Close Price (RMS) & Root mean square of closing prices &  & X &  & X \\ \midrule
Market Cap (Sum) & Total market capitalization sum &  & X & X & X \\ \midrule
Market Cap (Variance) & Variance of market capitalization &  & X &  &  \\ \midrule
Close Price Change (RMS) & RMS of closing price changes &  & X & X &  \\ \midrule
User Interactions (Min) & Minimum user interactions &  &  & X & X \\ \midrule
High Interaction Points (Sum) & Sum of high interaction points &  &  & X &  \\ \midrule
Sentiment Score (Sum) & Total sentiment score &  &  & X & X \\ \midrule
Sentiment Score (RMS) & RMS of sentiment scores &  &  & X &  \\ \midrule
Sentiment Score (Max) & Maximum sentiment score &  &  & X & X \\ \midrule
Sentiment Score (Min) & Minimum sentiment score &  &  & X & X \\ \midrule
High Sentiment Points (Sum) & Sum of high sentiment points &  &  & X &  \\ \midrule
Posts Created (Sum) & Total posts created &  &  & X &  \\ \midrule
Posts Created (Min) & Minimum posts created &  &  & X & X \\ \midrule
Social Dominance (Sum) & Total social dominance &  &  & X & X \\ \midrule
High Social Dominance (Sum) & High social dominance points &  &  & X & X \\ \midrule
Interaction-to-Market Cap Ratio (Min) & Min. interaction-to-market cap ratio &  &  & X & X \\ \midrule
Close Price Spectral Entropy (Sum) & Spectral entropy of closing prices &  &  &  & X \\ \midrule
Close Price Spectral Entropy (Median) & Median spectral entropy &  &  &  & X \\ \midrule
Close Price RMS (Sum) & RMS of closing prices (sum) &  &  &  & X \\ \midrule
Close Price RMS (Median) & Median RMS of closing prices &  &  &  & X \\ \midrule
Close Price Hurst Exponent (Sum) & Sum of Hurst exponents &  &  &  & X \\ \bottomrule
\end{tabular}%
}
\end{table}

\subsubsection{Hyperparameter Tuning}
After feature selection, we trained the XGBoost classifier on various feature combinations. Hyperparameters, including~\texttt{max\_depth},~\texttt{min\_child\_weight}, \\
and~\texttt{scale\_pos\_weight}, were fine-tuned using grid search with 3-fold cross-validation. Interestingly, the default parameters of the \texttt{XGBClassifier} in the Python \texttt{xgboost} library performed better than the grid-searched parameters.

\subsubsection{Performance Evaluation}
\begin{figure*}[h]
\centering
\includegraphics[width=0.7\textwidth]{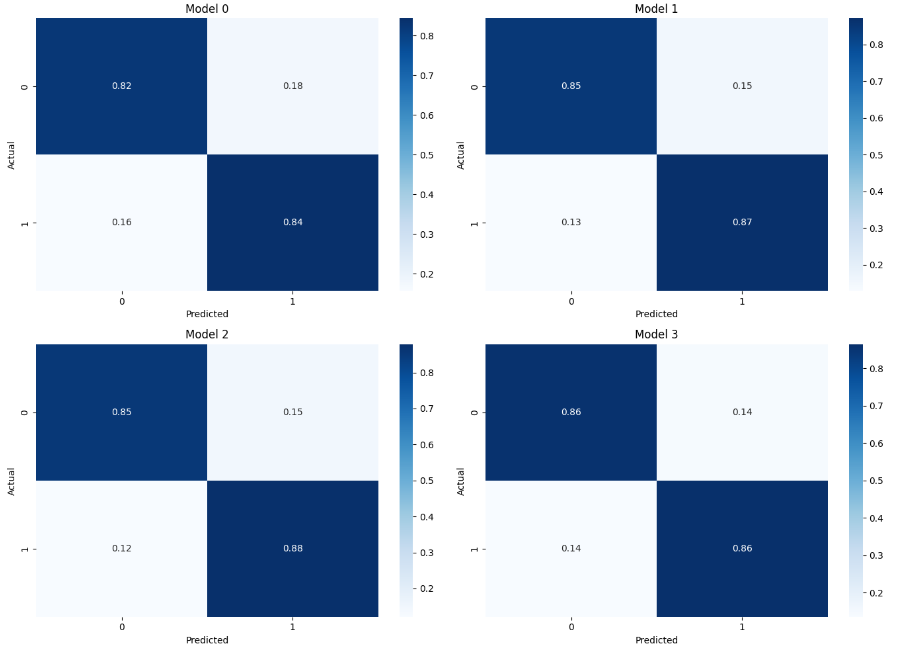}
\caption{Confusion Matrices of XGBoost Models for Different Features}
\label{fig:confusion_matrices}
\end{figure*}

\begin{figure}[t]
\centering
\includegraphics[width=\columnwidth]{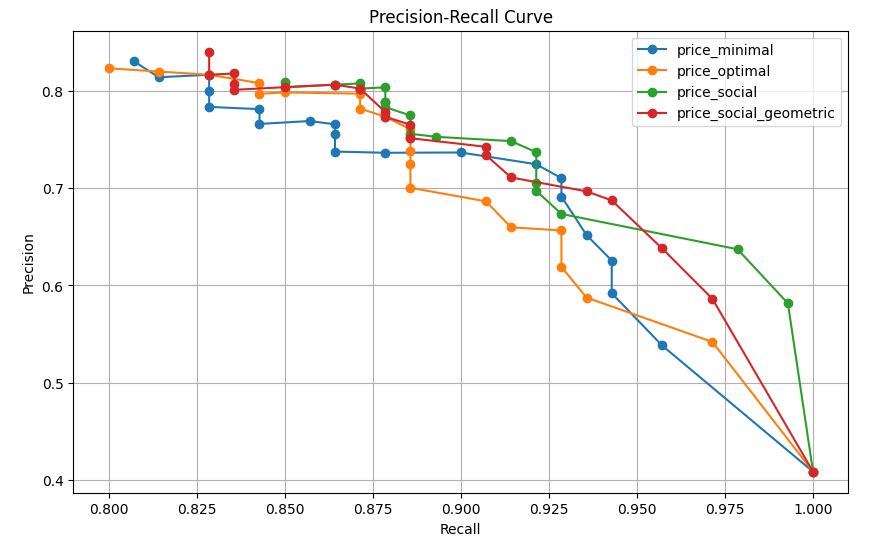}
\caption{Precision-Recall Curve for Models (Model 0: \textit{price\_minimal}, Model 1: \textit{price\_optimal}, Model 2: \textit{price\_social}, Model 3: \textit{price\_social\_geometric})}
\label{fig:f4}
\end{figure}

\begin{figure}[t]
\centering
\includegraphics[width=\columnwidth]{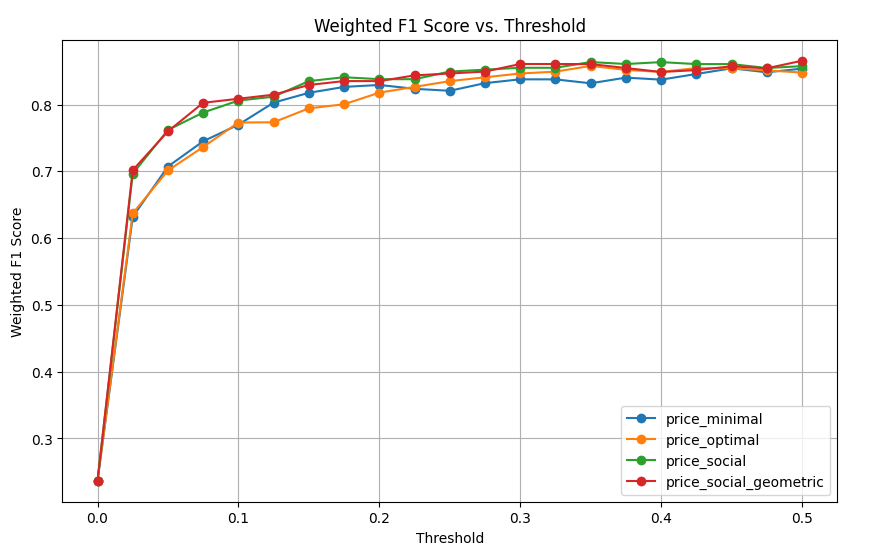}
\caption{Weighted F1 Score vs. Threshold for Models}
\label{fig:f5}
\end{figure}

\begin{figure}[t]
\centering
\includegraphics[width=\columnwidth]{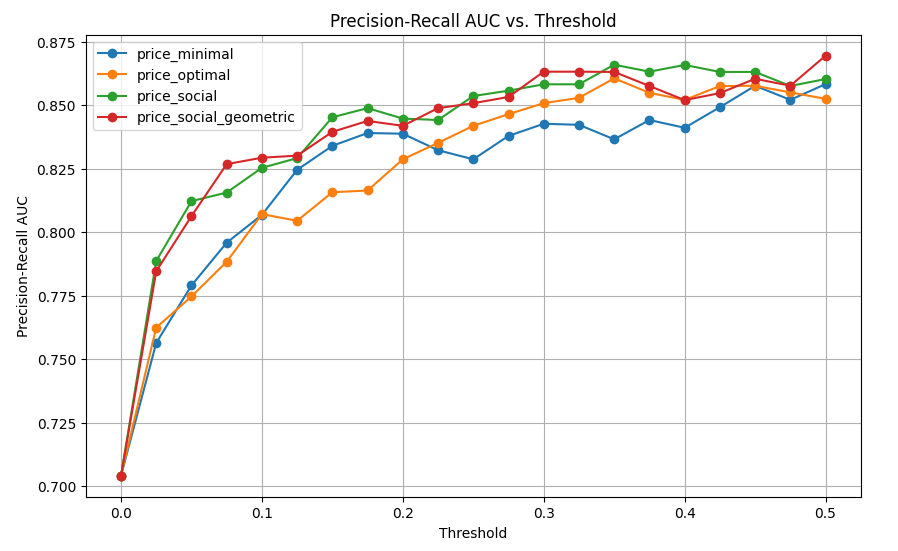}
\caption{Precision-Recall AUC vs. Threshold for Models}
\label{fig:f6}
\end{figure}

We evaluated performance using metrics like precision, recall, F1 score, and Precision-Recall AUC (Figures \ref{fig:f4}, \ref{fig:f5}, and \ref{fig:f6}). Table \ref{tab:key_features} highlights the reliance of all models on market data, such as maximum and minimum closing prices and market cap metrics. While social data (e.g., sentiment scores and user interactions) and geometric time series features (e.g., spectral entropy) marginally improved models 2 and 3, traditional market data remained the primary predictive factor.

\subsection{Unsupervised Clustering Experiments}
Finding potential coins that have a higher likelihood of being "pumped" in the future was a crucial phase of our research. We explored clustering, an unsupervised learning method frequently used to detect trends or anomalies in large datasets, such as cryptocurrency price data. Our analysis employed clustering algorithms, including Gaussian Mixture Models, K-means, Agglomerative Clustering, and DBSCAN. However, these methods did not produce strong results in identifying distinct groups within the data, likely due to the complexity and overlap of the dataset.

\subsubsection{Data Preprocessing}
\textbf{Feature Engineering:} The initial dataset included a variety of social and financial metrics extracted using the \texttt{tsfresh} library. Features were chosen based on their importance in detecting P\&D schemes.\\
\textbf{Log Transformation and Scaling:} Log-transformed features were used to reduce skewness and stabilize variance. To ensure equal contribution from each feature, data was standardized using \texttt{StandardScaler}.\\
\textbf{Principal Component Analysis (PCA):} PCA was applied to reduce dimensionality and preserve components capturing the greatest variance, optimizing the clustering procedure and aiding visualization of high-dimensional data.\\

\subsubsection{DBSCAN (Density-Based Spatial Clustering of Applications with Noise)}
DBSCAN groups points based on density while classifying isolated points as noise. Unlike other methods, it does not require specifying the number of clusters. Instead, it uses two parameters: \texttt{min\_samples}, the minimum points to form a dense region, and \texttt{eps}, the radius of each neighborhood \cite{ester1996density}.\\

\subsubsection{K-means Clustering}
K-means is widely used for dividing datasets into $k$ clusters. The algorithm iteratively updates cluster centers (centroids) and reassigns points until convergence \cite{macqueen1967kmeans}. In our study, K-means grouped coins based on historical data patterns. Multiple values of $k$ were tested to identify performance tiers among coins.\\

\subsubsection{Agglomerative Clustering}
Agglomerative clustering is a hierarchical method where each data point starts as its own cluster, merging iteratively based on a chosen linkage criterion \cite{murtagh2012agglomerative}. This method revealed multi-level hierarchies, helping visualize how clusters evolve and merge as distance thresholds increase.\\

\subsubsection{Gaussian Mixture Model (GMM)}
GMM is a probabilistic approach that assumes data originates from multiple Gaussian distributions. Parameters for each cluster are estimated using the Expectation-Maximization (EM) algorithm \cite{reynolds2009gmm}. GMM is particularly useful for handling clusters with varying shapes and sizes.\\

\begin{table}[htbp]
    \centering
    \caption{Clustering Algorithm Performance Metrics}
    \label{tab:f3}
    \begin{tabular}{l r r}
    \toprule
    \textbf{Clustering Algorithm} & \textbf{Silhouette Score} & \textbf{Davies-Bouldin Score} \\  
    \midrule
    DBSCAN & 0.19 & 1.55 \\
    K-means & 0.55 & 0.74 \\
    Agglomerative & 0.46 & 0.81 \\
    Gaussian Mixture (GMM) & 0.48 & 0.78 \\
    \bottomrule
    \end{tabular}
\end{table}

\begin{figure*}[tb]
    \centering
    \includegraphics[width=0.8\textwidth]{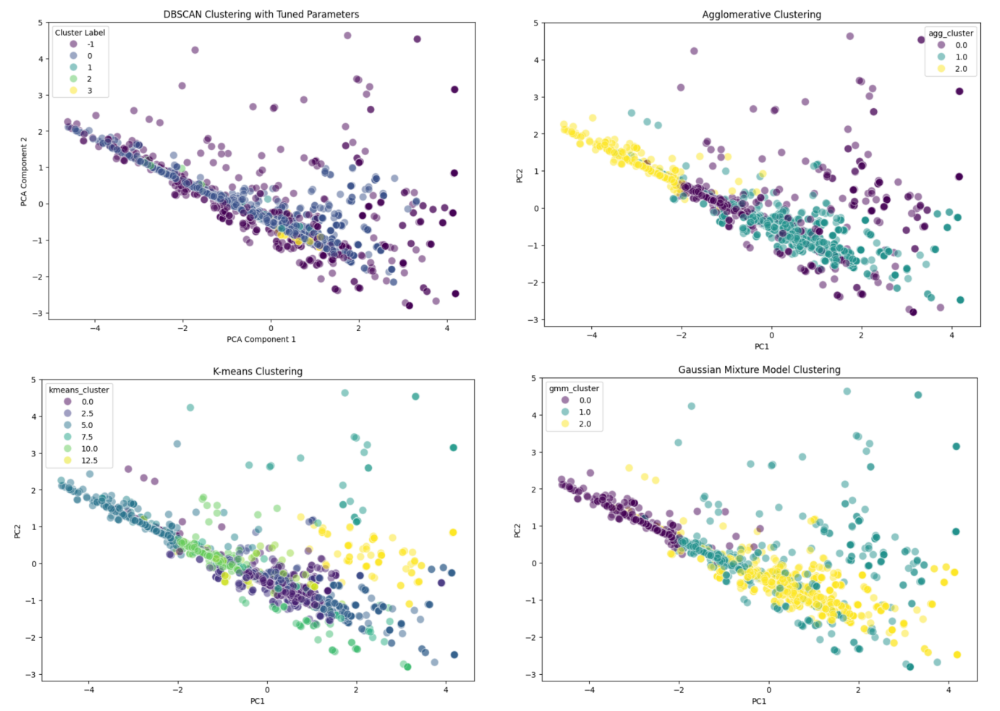}
    \caption{Comparison of Different Clustering Methods}
    \label{fig:clustering}
\end{figure*}

As shown in Table \ref{tab:f3}, the Silhouette and Davies-Bouldin scores indicate poor clustering performance. Low Silhouette scores suggest that clusters are not well-defined, while high Davies-Bouldin scores imply significant overlap among clusters. Despite extensive preprocessing, the dataset's inherent structure likely contributes to these challenges. The characteristics of the clusters appear insufficiently distinct, complicating the ability of algorithms to differentiate groups effectively.



\section{Additional Figures and Tables}

\begin{table}[H]
\centering
\small
\caption{Number of P\&D events by exchange.}
\label{tab:pumps_by_exchange}
\begin{tabular}{llrr}
    \toprule
    \textbf{Exchange} & \textbf{Type} & \textbf{Pumps} & \textbf{Percentage} \\
    \midrule
    Hotbit & CEX & 324 & 15.6\% \\
    LATOKEN & CEX & 304 & 14.6\% \\
    XT & CEX & 268 & 12.9\% \\
    Pancakeswap & DEX & 247 & 11.9\% \\
    Poloniex & CEX & 246 & 11.8\% \\
    KuCoin & CEX & 181 & 8.7\% \\
    LBank & CEX & 150 & 7.2\% \\
    DigiFinex & CEX & 120 & 5.8\% \\
    MEXC & CEX & 72 & 3.5\% \\
    Binance & CEX & 45 & 2.2\% \\
    Others & Mixed & 122 & 5.8\% \\
    \bottomrule
\end{tabular}
\end{table}

\begin{table}[H]
\centering
\small
\caption{Top five pumped coins and their token standards.}
\label{tab:top_5}
\begin{tabular}{lllr}
    \toprule
    \textbf{Symbol} & \textbf{Name} & \textbf{Standard} & \textbf{Pumps} \\
    \midrule
    TOKKI & CRYPTOKKI & ERC-20 & 9 \\
    DXGM & DexGame & ERC-20 & 8 \\
    HMR & Homeros Game Barracks & ERC-20 & 8 \\
    NAR & Narwhalswap & BEP-20 & 7 \\
    JUSTICE & AssangeDAO & ERC-20 & 7 \\
    \bottomrule
\end{tabular}
\end{table}

\begin{figure}[H]
    \centering
    \includegraphics[width=0.45\textwidth]{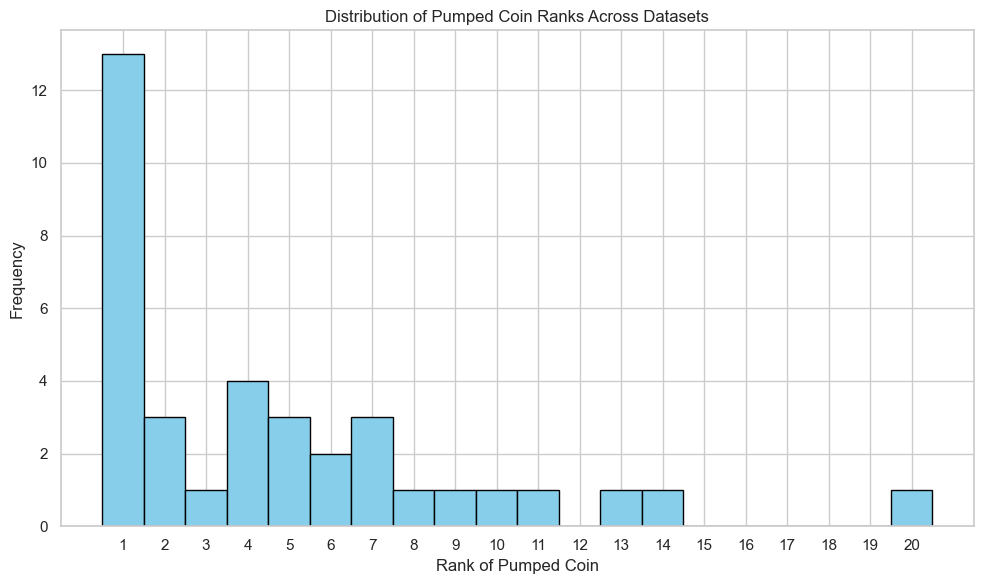}
    \caption{Distribution of pumped coin ranks across P\&D events.}
    \label{fig:histogram_rank_pummps}
\end{figure}

\begin{figure*}[b]
    \centering
    \begin{minipage}[t]{0.4\textwidth}
        \centering
        \includegraphics[width=\textwidth]{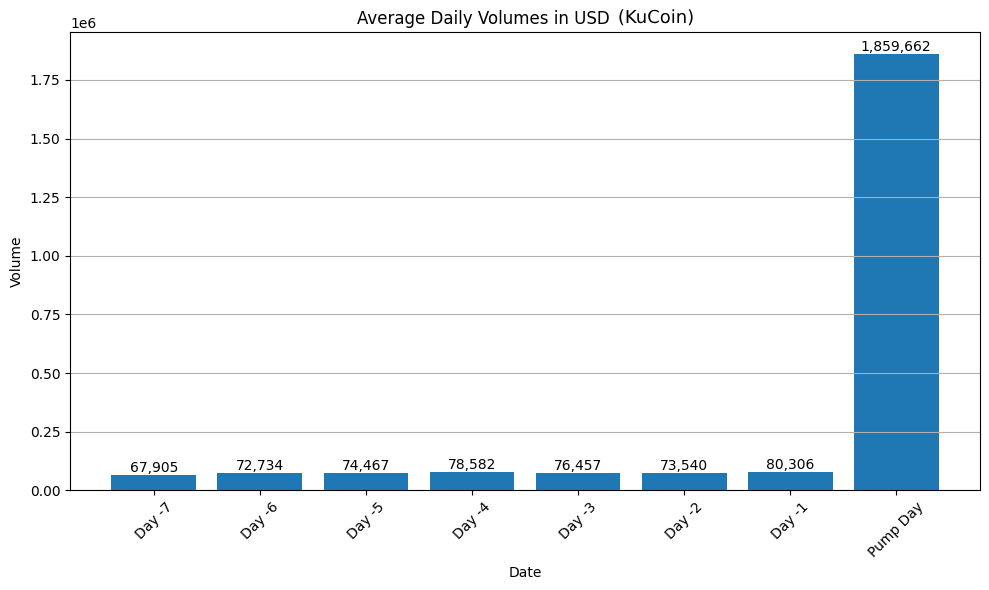}
        \caption{Average Daily Trade Volumes on KuCoin.}
        \label{fig:avg_daily_volumes_kucoin}
    \end{minipage}%
    \hfill
    \begin{minipage}[t]{0.4\textwidth}
        \centering
        \includegraphics[width=\textwidth]{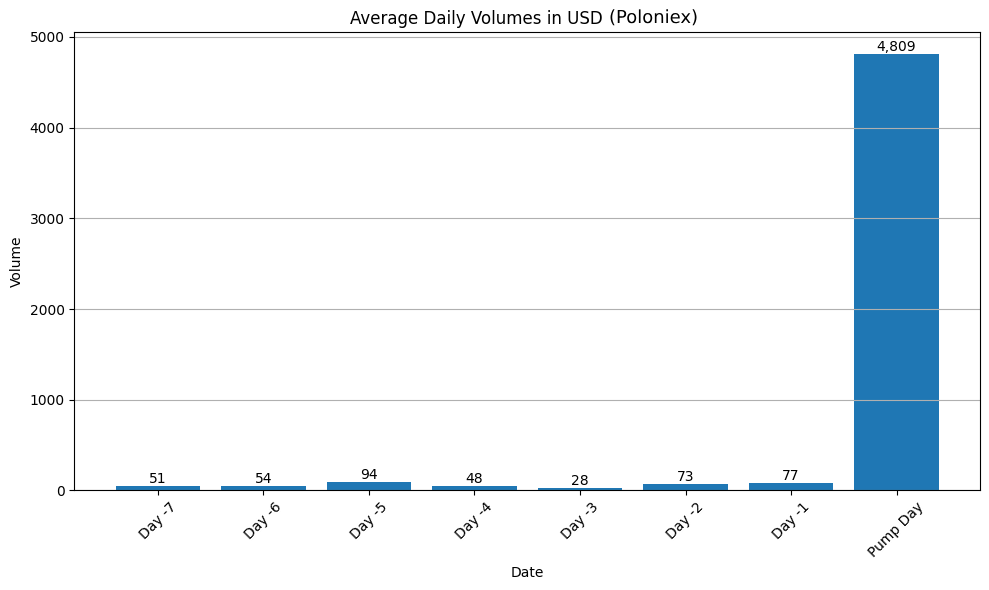}
        \caption{Average Daily Trade Volumes on Poloniex.}
        \label{fig:avg_daily_volumes_poloniex}
    \end{minipage}
    
    \vspace{0.5cm}
    
    \begin{minipage}[t]{0.4\textwidth}
        \centering
        \includegraphics[width=\textwidth]{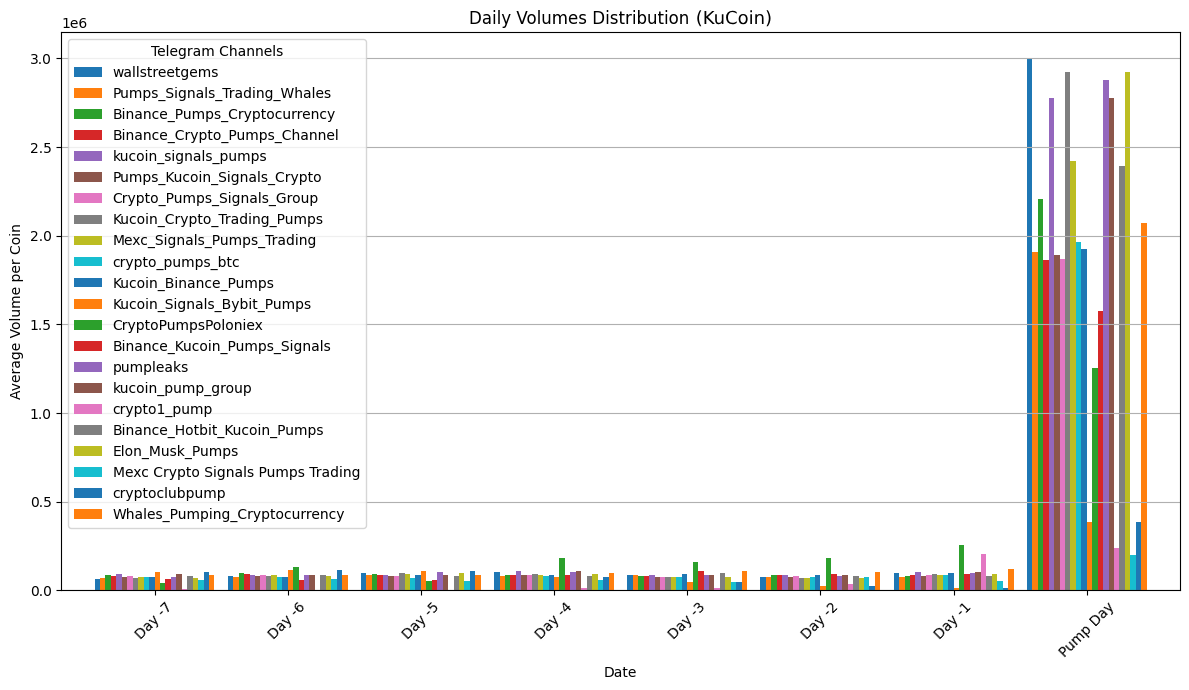}
        \caption{Average Daily Trade Volumes on KuCoin by Telegram Channel.}
        \label{fig:avg_daily_volumes_by_channels_kucoin}
    \end{minipage}%
    \hfill
    \begin{minipage}[t]{0.4\textwidth}
        \centering
        \includegraphics[width=\textwidth]{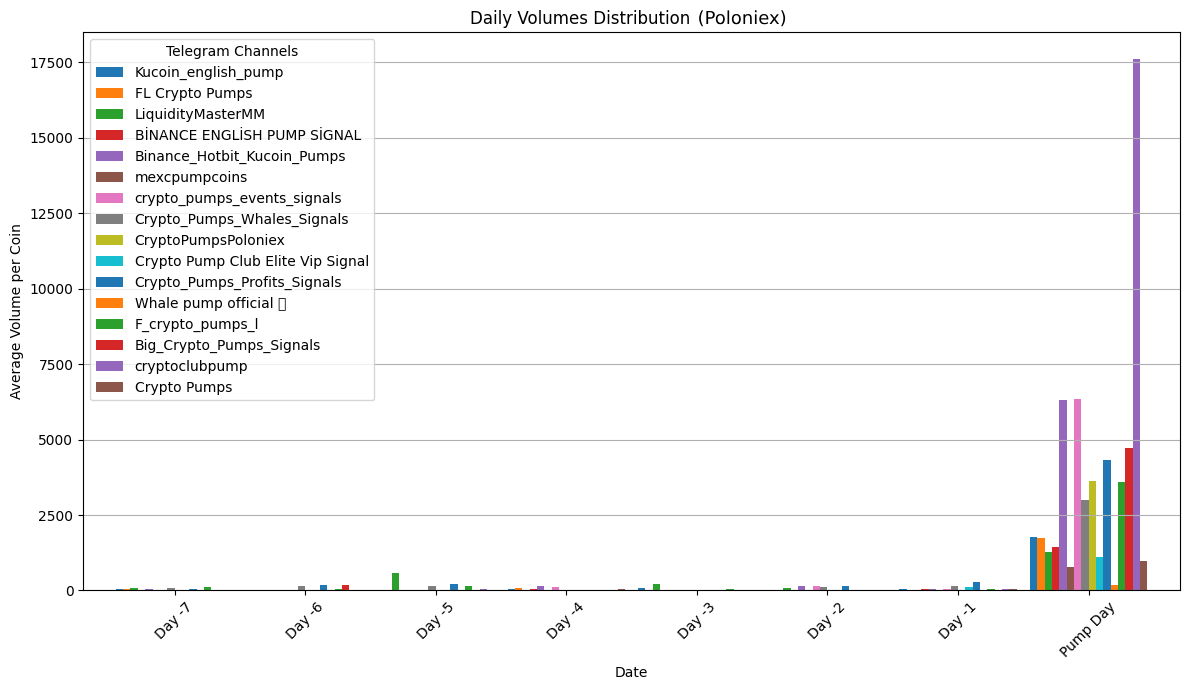}
        \caption{Average Daily Trade Volumes on Poloniex by Telegram Channel.}
        \label{fig:avg_daily_volumes_by_channels_poloniex}
    \end{minipage}
    
    \vspace{0.5cm}
    
    \begin{minipage}[t]{0.4\textwidth}
        \centering
        \includegraphics[width=\textwidth]{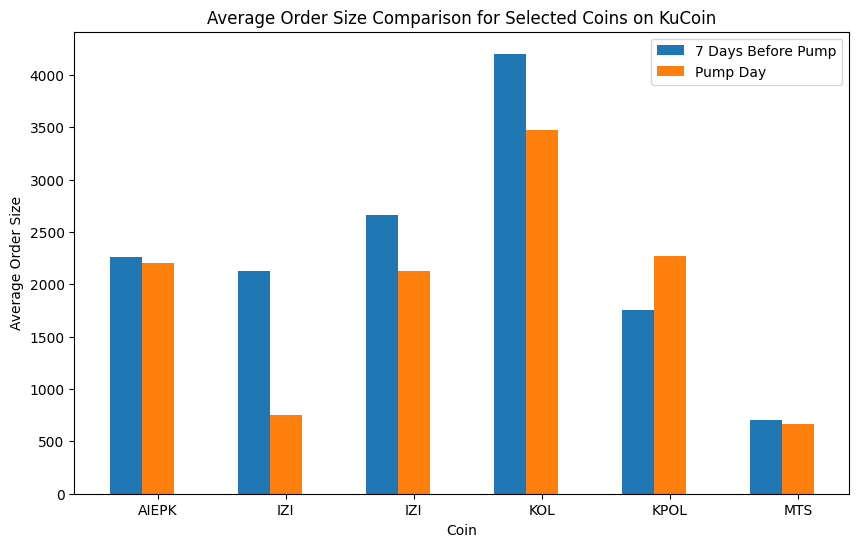}
        \caption{Average Order Size Comparison for KuCoin.}
        \label{fig:kucoin_bar_chart}
    \end{minipage}%
    \hfill
    \begin{minipage}[t]{0.4\textwidth}
        \centering
        \includegraphics[width=\textwidth]{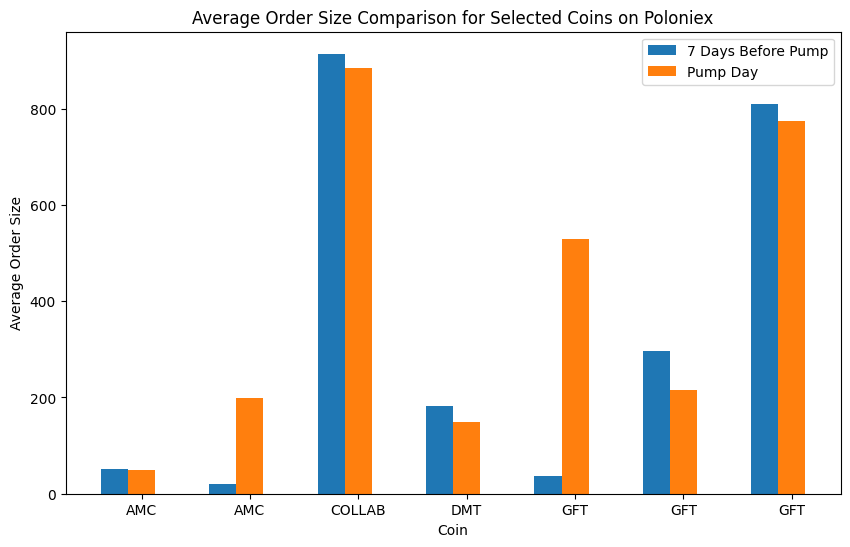}
        \caption{Average Order Size Comparison for Poloniex.}
        \label{fig:poloniex_order_size}
    \end{minipage}
\end{figure*}

\end{document}